\documentclass[12pt]{article}
\usepackage{graphicx}
\usepackage[toc,page]{appendix}
\def\dlim{\mathrel{\raise.8ex\hbox{${\scriptscriptstyle D =
4}$\kern-1.5em\lower1ex\hbox{$\longrightarrow$}}}}
\def\ltwid{\mathrel{\raise.3ex\hbox{$<$\kern-.75em\lower1ex\hbox{$\sim$}}}}
\def \be{\begin{equation}}
\def \ee{\end{equation}}
\def \bea{\begin{eqnarray}}
\def \eea{\end{eqnarray}}
\def\fint{\rlap{$\rfloor$}\lceil}
\def\Fint{\rlap{$\Biggl\rfloor$}\Biggl\lceil}
\def\ltwid{\mathrel{\raise.3ex\hbox{$<$\kern-.75em\lower1ex\hbox{$\sim$}}}}
\def\gtwid{\mathrel{\raise.3ex\hbox{$>$\kern-.75em\lower1ex\hbox{$\sim$}}}}

\def\square{\kern1pt\vbox{\hrule height 1.2pt\hbox{\vrule width 1.2pt\hskip 3pt
   \vbox{\vskip 6pt}\hskip 3pt\vrule width 0.6pt}\hrule height 0.6pt}\kern1pt}

\def \p{\prime}

\begin{document}

\begin{titlepage}

\begin{flushright}
ITP-UU-13/13, SPIN-13/09 \\ IGC-13/6-1
\end{flushright}

\vspace{2cm}

\begin{center}
{\bf Alternate Definitions of Loop Corrections to the
Primordial Power Spectra}
\end{center}

\vspace{.5cm}

\begin{center}
S. P. Miao$^*$
\end{center}

\begin{center}
\it{Institute for Theoretical Physics, Spinoza Institute, University of Utrecht
\\Luevenlaan 4, Postbus 80.195, 3508TD Utrecht, NETHERLANDS }
\it{Department of Physics, National Cheng Kung University
\\ No. 1, University Road, Tainan 701, Taiwan}
\end{center}

\begin{center}
Sohyun Park$^{\dagger}$
\end{center}

\begin{center}
\it{Institute for Gravitation and the Cosmos, The Pennsylvania State
University, University Park, PA 16802, UNITED STATES}
\end{center}

\begin{center}
ABSTRACT
\end{center}
We consider two different definitions for loop corrections
to the primordial power spectra. One of these is to simply correct
the mode functions in the tree order relations using the linearized
effective field equations. The second definition involves the spatial
Fourier transform of the 2-point correlator. Although the two
definitions agree at tree order, we show that they disagree at one
loop using the Schwinger-Keldysh formalism, so there are at least two
plausible ways of loop correcting the tree order result. We discuss
the advantages and disadvantages of each.

\vspace{2cm}

\begin{flushleft}
\end{flushleft}

\begin{flushleft}
$^*$ e-mail: S.Miao@uu.nl ; spmiao5@mail.ncku.edu.tw\\
$^{\dagger}$ e-mail: spark@gravity.psu.edu
\end{flushleft}

\end{titlepage}

\section{Introduction}

It is clear that
the tensor \cite{Starobinsky} and scalar \cite{Mukhanov} power
spectra from primordial inflation are quantum gravitational effects
by how the approximate tree order results depend upon Planck's
constant $\hbar$ and Newton's constant $G$,
\begin{equation}
\Delta^2_{h}(k) \approx \frac{16 \hbar G H^2(t_k)}{\pi c^5} \qquad ,
\qquad \Delta^2_{\mathcal{R}}(k) \approx \frac{\hbar G H^2(t_k)}{\pi
c^5 \epsilon(t_k)} \; . \label{Hubbleform}
\end{equation}
(Here $H(t)$ is the Hubble parameter, $\epsilon(t)$
is the first slow roll parameter, and $t_k$ is the time of
first horizon crossing for the mode of wave number $k$\footnote{
The definition of $t_{k}$ is the time at which the physical
wave number $k/a(t)$ of some perturbation equals the Hubble parameter,
$k = H(t_k) a(t_k)$.}.) These effects were predicted
around 1980, and the detection of the scalar power spectrum in 1992
\cite{COBE} represents the first quantum gravitational data ever taken.
Much more has followed \cite{WMAP,PLANCK}, as have increasingly sensitive
bounds on the tensor power spectrum \cite{SPT,ACP}.
Although using this data to study quantum gravity has so far been limited
by the absence of a compelling model for inflation,
there is no objection to the revolutionary character of these events.

The tree order results (\ref{Hubbleform}) are just the first terms
in the quantum loop expansion in which each higher loop is suppressed
by an additional factor of $G H^2$.
Assuming single-scalar inflation, the best
current data bounds this loop-counting parameter to be no larger than about
$G H^2 \ltwid 10^{-10}$ \cite{KOW}. That is a very small number, but it has
been suggested that the sensitivity to resolve one loop corrections might be
obtained by measuring the matter power spectrum out to redshifts of as high
as $z \sim 50$ \cite{21cm}. Reaching that goal would be very difficult, requiring
both a unique model of inflation to pin down the tree order contribution and
a secure understanding of the relevant astrophysics to extract the primordial
signal. However, the work is in progress \cite{first21}, and the project does not
seem hopeless.

The possibility of resolving one loop corrections to the power spectra
has motivated theorists to do intensive studies on the issue
\cite{Weinberg, others}. Because the effect will necessarily be very small,
much attention has been devoted to potentially large enhancements from factors
of $1/\epsilon$ in the $\zeta$ propagator \cite{SLS,JaSl,XGB}, and from
the formal infrared divergence \cite{IRstudies} of $\zeta$ and
graviton propagators implied by the approximate scale invariance of their
tree order power spectra (\ref{Hubbleform}). This has raised the issue of
precisely defining {\it what} is being loop corrected. The tree order results
(\ref{Hubbleform}) are consistent with the spatial Fourier transform of
2-point correlators of the graviton and $\zeta$ fields,
\begin{eqnarray}
\Delta^2_{h}(k) & \equiv & \lim_{t \gg t_k} \frac{k^3}{2 \pi^2} \int
\!\! d^3x \, e^{-i\vec{k} \cdot \vec{x}} \Bigl\langle \Omega
\Bigl\vert h_{ij}(t,\vec{x}) h_{ij}(t,\vec{0}) \Bigr\vert \Omega
\Bigr\rangle \; , \label{tenVEV} \\
\Delta^2_{\mathcal{R}}(k) & \equiv & \lim_{t \gg t_k} \frac{k^3}{2
\pi^2} \int \!\! d^3x \, e^{-i\vec{k} \cdot \vec{x}} \Bigl\langle
\Omega \Bigl\vert \zeta(t,\vec{x}) \zeta(t,\vec{0}) \Bigr\vert
\Omega \Bigr\rangle \; . \label{scalVEV}
\end{eqnarray}
There is no question that one loop corrections to these expressions
show sensitivity to the infrared cutoff \cite{GidSloth}. This
sensitivity can be canceled by re-defining the ``power spectra'' as
the expectation values of appropriately chosen operators
\cite{gauge}. However, such redefinitions tend to alter the
$\epsilon$ dependence of loop corrections, and they also introduce
new ultraviolet divergences whose renormalization is not currently
understood \cite{issues}.

The point of this paper is to consider another generalization of
what is meant by the ``primordial power spectra.'' This alternate
generalization is motivated by the relations which emerge from
expressions (\ref{tenVEV}) and (\ref{scalVEV}) when one uses the
free field mode sums for $h_{ij}$ and $\zeta$,
\begin{eqnarray}
\Delta^2_{h}(k) & = & \lim_{t \gg t_k} \frac{k^3}{2 \pi^2} \times 64
\pi G \times \vert u(t,k) \vert^2 \; , \label{tenmode} \\
\Delta^2_{\mathcal{R}}(k) & = & \lim_{t \gg t_k} \frac{k^3}{2 \pi^2}
\times 4 \pi G \times \vert v(t,k) \vert^2 \; , \label{scalmode}
\end{eqnarray}
where $u(t,k)$ and $v(t,k)$ are the plane wave mode functions\footnote
{$u(t,k)$ and $v(t,k)$ are not the one-particle-irreducible(1PI)
1-point functions of $h_{ij}(t, \vec{x})$ and $\zeta(t,\vec{x})$
respectively.} of tensor and scalar perturbations. The alternate generalization
is to simply extend the tree order relations (\ref{tenmode})
and (\ref{scalmode}) to all orders using the mode functions obtained
by solving the linearized Schwinger-Keldysh effective field equations\footnote{
A curious reader might wonder how the quantum corrected mode functions are related
to the Heisenberg operators which satisfy the standard commutation relations.
We demonstrate the relation between them using our worked-out example
in Appendix \ref{cano-mode}.}.
Even though equations (\ref{tenVEV}, \ref{scalVEV}) and
(\ref{tenmode}, \ref{scalmode}) are two different approaches of
quantum correcting primordial power spectra,
the diagram topology for quantum corrections to the mode function definition
is identical to that of quantum corrections to the correlator\footnote{
The generic diagram topology for the two definitions is derived in Appendix \ref{topology}.}.
Also, note that the two definitions would agree if the in-out formalism
had been employed. However, one must employ the Schwinger-Keldysh formalism
in cosmological scenarios. It is not so clear whether or not the two definitions agree
at one loop due to subtle differences in which of the four Schwinger-Keldysh propagators
appears. That is what we shall check.

In this paper we start with briefly reviewing single scalar inflation,
deriving the tree order results and reasoning alternate definitions.
This comprises of section 2. In section 3 we digress to sketch the
Schwinger-Keldysh formalism\footnote{It is also called the in-in or the
closed time path formalism.} and give rules to facilitate our computation.
In section 4 we use a worked-out example to demonstrate that two definitions
disagree at one loop. Finally we discuss the advantages and disadvantages
for each definition in section 5.

\section{Two alternate definitions for loop-corrected primordial power spectra}
Primordial power spectra not only allow us to understand the early Universe,
but also serve as a bridge that connects cosmology with fundamental theory. For
example, resolving the tensor power spectrum would confirm the existence
of gravitons and their quantization. Attaining the sensitivity to resolve
loop corrections to the power spectra would, along with a unique theory of inflation,
direct theorists in the construction of a renormalizable theory of quantum gravity.

Two of the many frustrations in the attempt to connect inflation with fundamental
theory are first, we lack a unique model of inflation  --- which means we don't know
the time dependence of the scale factor $a(t)$ --- and
second, we do not have a solution for
the tree order mode functions for a general $a(t)$ even if we happened to know it.
This means that approximations must be used even for the tree order power spectra. It
also implies that we must approximate the propagators which occur in loop integrations
because these propagators are mode sums of products of unknown tree order mode
functions. These are all important problems, but here we wish to focus on
the issue of what theoretical quantity represents the observed power spectrum.
That is, what quantity would we like to compute, assuming we had the mode functions
and propagators necessary to make the computation? In particular, is it the spatial
Fourier transform of the 2-point correlators (\ref{tenVEV}) and (\ref{scalVEV}), or
should we instead use the norm squared of the mode functions (\ref{tenmode}) and
(\ref{scalmode})? We begin with a quick review of single scalar inflation which
is meant to pedagogically demonstrate that the two definitions coincide at tree order.
The burden is that they disagree at one loop.

The dynamical variables of single-scalar inflation are
the metric ${\rm g}_{\mu\nu}(t,\vec{x})$ and the inflaton field $\varphi(t,\vec{x})$.
Its Lagrangian density is,
\begin{equation}
\mathcal{L} = \frac1{16 \pi G} \, {\rm R} \sqrt{-{\rm g}} -\frac12
\partial_{\mu} \varphi \partial_{\nu} \varphi {\rm g}^{\mu\nu}
\sqrt{-{\rm g}} - V(\varphi) \sqrt{-{\rm g}} \; . \label{Ldef}
\end{equation}
Primordial inflation can be described by homogeneous, isotropic
and spatially flat background metric,
\begin{equation}
{\rm g}^0_{\mu\nu} dx^{\mu} dx^{\nu} = -dt^2 + a^2(t) d\vec{x}
\!\cdot\! d\vec{x} \; ,
\end{equation}
with the slow roll parameter,
\begin{equation}
\epsilon(t) \equiv-\frac{\dot{H}}{H^2} \qquad , \qquad
0<\epsilon(t)<1\; .\label{slowroll}
\end{equation}
Here $H(t)$ is the Hubble parameter defined as the first time derivative
of the scale factor $a(t)$,
\begin{equation}
H(t) \equiv \frac{\dot{a}}{a}.
\end{equation}
It indicates whether or not the Universe is expanding.

We follow the convention of Maldacena \cite{JM} and Weinberg \cite{SW}
for decomposing the spatial metric\footnote{This spatial metric $g_{ij}$
is from Arnowitt-Deser-Misner (ADM) decomposition
\cite{ADM}:
${\rm g}_{00} \equiv -N^2 \!+\! g_{ij} N^i N^j \;,\;
{\rm g}_{0i} \equiv -g_{ij} N^j \:,\: {\rm g}_{ij} \equiv g_{ij}\:.$ },
\begin{eqnarray}
&&g_{ij}(t,\vec{x}) \equiv a^2(t) e^{2 \zeta(t,\vec{x})}
\widetilde{g}_{ij}(t,\vec{x}) \; \\
&&\widetilde{g}_{ij}(t,\vec{x}) \equiv \Bigl( e^{h(t,\vec{x})}
\Bigr)_{ij} = \delta_{ij} + h_{ij} + \frac12 h_{ik} h_{kj} + \dots
\label{hdef}\;,
\end{eqnarray}
where the $\zeta(t,\vec{x})$ and $h_{ij}(t,\vec{x})$ fields
are the scalar and tensor perturbations respectively.
During the $50$ e-foldings of primordial inflation which is required
to explain the horizon problem, many modes must experience first horizon
crossing, $k=a(t_k)H(t_k)$. After that time they became almost constant
and survived to be detected today.
Therefore the tensor and scalar power spectra are defined (for $D = 4$
spacetime dimensions) as in (\ref{tenVEV}) and (\ref{scalVEV}).

Maldacena \cite{JM} and Weinberg \cite{SW} employ Arnowitt-Deser-Misner
(ADM) notation but they do not fix the gauge by specifying lapse $N(t,\vec{x})$
and shift $N^i(t,\vec{x})$ functions. They instead fix the surface of simultaneity
using the background value of the inflaton, $\varphi(t,\vec{x})=\varphi_0(t)$,
and impose the spatial transverse gauge condition, $\partial_j h_{ij}(t,\vec{x}) = 0$.
The lapse and shift functions hence\footnote{$N[\zeta,h](t,\vec{x})$ can be solved
exactly \cite {KOW} but there only exists a perturbative solution for
$N^i[\zeta,h](t,\vec{x})$.} can be determined as nonlocal functionals of graviton fields
from solving the gauged fixed constraint equations. Substituting those solutions into
the original Lagrangian, it is not so hard to obtain the quadratic part,
\begin{eqnarray}
\mathcal{L}_{h^2} & = & \frac{a^{D-1}}{64\pi G} \Biggl\{
\dot{h}_{ij} \dot{h}_{ij} - \frac1{a^2} \partial_k h_{ij} \partial_k
h_{ij} \Biggr\} \;, \label{freeh}\\
\mathcal{L}_{\zeta^2} & = & \frac{(D \!-\!2) \, \epsilon \,
a^{D-1}}{ 16\pi G} \Biggl\{ \dot{\zeta}^2 - \frac1{a^2} \partial_k
\zeta \partial_k \zeta \Biggr\} \; . \label{freezeta}
\end{eqnarray}
From expression (\ref{freeh}) we see that each of the $\frac12 (D-3) D$
graviton polarizations is $\sqrt{32 \pi G}$ times a canonically
normalized, massless, minimally coupled scalar. Its plane wave mode
function $u(t,k)$ obeys,
\begin{equation}
\ddot{u} + (D \!-\! 1) H \dot{u} + \frac{k^2}{a^2} u = 0 \qquad {\rm
with} \qquad u \dot{u}^* - \dot{u} u^* = \frac{i}{a^{D-1}} \; .
\label{hmodes}
\end{equation}
Expression (\ref{freezeta}) implies that the free field expansion for
$\zeta(t,\vec{x})$ is $\sqrt{8\pi G/(D-2)}$ times a canonically
normalized scalar whose plane wave mode functions $v(t,k)$ obey,
\begin{equation}
\ddot{v} + \Bigl[(D \!-\! 1) H \!+\!
\frac{\dot{\epsilon}}{\epsilon} \Bigr] \dot{v} +
\frac{k^2}{a^2} v = 0 \qquad {\rm with} \qquad
v\dot{v}^* - \dot{v}v^* =\frac{i}{\epsilon a^{D-1}} \; .
\label{zetamodes}
\end{equation}
To derive equations (\ref{tenmode}) and (\ref{scalmode}) (in $D=4$ spacetime dimensions)
for the primordial power spectra one substitutes the free field expansions
for $h_{ij}(t,\vec{x})$ and $\zeta(t,\vec{x})$ into equations (\ref{tenVEV})
and (\ref{scalVEV}).

From the tree order derivation for the tensor power spectrum we establish
the following relation\footnote{The relation for the scalar power spectrum
reaches the same form.},
\begin{eqnarray}
\frac{k^3}{2\pi^2}\lim_{t\geq t_k}\Biggl\{\int\!\! d^3 x
e^{-i\vec{k}\cdot\vec{x}}
\Bigl\langle\Omega\Bigl\vert h_0(t,\vec{x})h_0(t,0)
\Bigr\vert\Omega\Bigr\rangle = \sharp \Bigl\vert u(t,k)\Bigr\vert^2
\Biggr\}\;,\label{treeid}
\end{eqnarray}
here we suppress tensor indexes and $\sharp$ is a constant which depends upon
the field we consider. Each side of equation (\ref{treeid}) has a clear
generalization to higher orders,
\begin{eqnarray}
&&\hspace{-1cm}\bullet\int\!\! d^3 x e^{-i\vec{k}\cdot\vec{x}}
\Bigl\langle\Omega\Bigl\vert h_0(t,\vec{x})h_0(t,0)
\Bigr\vert\Omega\Bigr\rangle\longrightarrow
\int\!\! d^3 x e^{-i\vec{k}\cdot\vec{x}}
\Bigl\langle\Omega\Bigl\vert h(t,x)h(t,0)
\Bigr\vert\Omega\Bigr\rangle; \label{VEVdef}\\
&&\hspace{-1cm}\bullet\;\sharp \Biggl\vert u_{0}(t,k)\Biggr\vert^2
\longrightarrow
\sharp \Biggl\vert\;u(t,k)\!+\!\sum_{l=1}\!
\Delta u_{l}(t,k)\Biggr\vert^2\nonumber\\
&&\hspace{1.2cm}=\sharp \Biggl\{\Bigl\vert u(t,k)\Bigr\vert^2
\!+\!\Delta u_1(t,k)u^*(t,k)
\!+\!\Delta u^*_1(t,k)u(t,k)\!+\!\cdots\Biggr\},
\label{modedef}
\end{eqnarray}
where higher order mode functions can be solved by the linearized
Schwinger-Keldysh effective field equation\footnote{
$\Delta u_0(t',k)\equiv u(t',k)$},
\begin{eqnarray}
\mathcal{D}\Bigl[\Delta u_l(t,k)e^{i\vec{k}\cdot\vec{x}}\;\Bigr]
\!=\!\!\int\!\! d^4 x' \sum_{k=1}^{l}\Bigl\{M^2_{\scriptscriptstyle ++}(x;x')\!
+M^2_{\scriptscriptstyle +-}(x;x')\Bigr\}_{k}\Delta u_{l-k}(t',k)
e^{i\vec{k}\cdot\vec{x'}}\;.\label{SWeqn}
\end{eqnarray}
Here $\mathcal{D}$ is
the kinetic operator.
Note that
$\frac{k^3}{2\pi^2}\lim_{t\geq t_k}$ in (\ref{treeid}) is a common factor
for both definitions. To simplify later discussion we drop it without
changing the generic structure of the two definitions.
At this step it is clear that one could compute the loop-corrected power spectra
either by spatially Fourier transforming the 2-point corrector --(\ref{VEVdef})
or exploiting the mode function definition --(\ref{modedef}).

\section{Schwinger-Keldysh formalism}
The purpose of this section is to give the rules for the various
Schwinger-Keldysh vertices and propagators. We also introduce the linearized
Schwinger-Keldysh effective field equation and demonstrate that
a causal result in $\varphi^3$ theory can be obtained by exploiting these rules.

For most of the problems we encounter in elementary particle physics
we are allowed to assume that quantum fields begin in free vacuum
at asymptotically early times and end up the same way at asymptotically
late times, for example, scattering processes in flat space.
However, this is not valid for cosmological settings in which
the in vacuum doesn't evolve to the out vacuum. The use of the
in-out formalism would result in quantum correction terms dominated
by events from the infinite future!
A realistic scenario corresponding to what we measure would rather be
that the Universe is released from a prepared state at a finite time and
allowed to evolve as it will. The Schwinger-Keldysh formalism can give
a correct description of this. Employing it \cite{JS,KTM,BM,LVK,CSHY,RDJ,CH,FW}
also guarantees that the computation is both real and causal.

It is convenient to sketch the in-in formalism by employing a scalar field
$\varphi(x)$. The basic construction is to evolve fields forwards with
$\fint[d\varphi_+]e^{S[\varphi_+]}$ from the time $i$ to the time $f$ and
backwards with $\fint[d\varphi_{-}]e^{S[\varphi_{-}]}$. To avoid a lengthy
digression, we give the key relation between the canonical operator and the
functional integral \cite{Miao,FW,TOW},
\begin{eqnarray}
\lefteqn{\Bigl\langle \Psi \Bigl\vert
\overline{T}^*\Bigl(\mathcal{O}_2[ \varphi]\Bigr)
T^*\Bigl(\mathcal{O}_1[\varphi]\Bigr) \Bigr\vert \Psi \Bigr\rangle =
\Fint [d\varphi_+] [d\varphi_-] \,
\delta\Bigl[\varphi_-(f) \!-\! \varphi_+(f)\Bigr] } \nonumber \\
& & \hspace{1.5cm} \times \mathcal{O}_2[\varphi_-]
\mathcal{O}_1[\varphi_+] \Psi^*[\varphi_-(i)] e^{i \int_{i}^{f} dt
\Bigl\{L[\varphi_+(t)] - L[\varphi_-(t)]\Bigr\}} \Psi[\varphi_+(i)]
\; ,\qquad \label{fund}
\end{eqnarray}
where $T^*$ stands for a time-ording symbol, except that
any derivatives are taken {\it outside} the time ordering, whereas
$\overline{T}^*$ is anti-time-ordered. Based on the same field
in (\ref{fund}) being represented by two different dummy functional
variables, $\varphi_{\pm}(x)$, several modified Feynman rules can be
inferred,
\begin{itemize}
\item{Each line has a polarity of either $+$ or $-$;}
\item{Vertices (and counterterms) are either all $+$ or all $-$;}
\item{Vertices (and counterterms) with $+$ polarity are the same as for the
usual Feynman rules and those with $-$ polarity have an extra minus sign;}
\item{External lines from the time-ordered operator carry $+$ polarity and
those from the anti-time-ordered operator carry $-$ polarity;}
\item{Propagators can be $++$, $-+$, $+-$ and $--$.}
\end{itemize}

Note also that we can directly read off the four propagators
from substituting the free Lagrangian in place of the full Lagrangian in
expression (\ref{fund}),
\begin{eqnarray}
i\Delta_{\scriptscriptstyle ++}(x;x^{\p})\!\!\! &=&\!\! \!
\Bigl\langle \Omega\Bigl\vert T\Bigl(\varphi(x) \varphi(x^{\p}) \Bigr)
\Bigr\vert\Omega \Bigr\rangle_0\; , \label{++}\\
i\Delta_{\scriptscriptstyle -+}(x;x^{\p})\!\!\!& =&\!\!\!
\Bigl\langle \Omega \Bigl\vert\varphi(x) \varphi(x^{\p})
\Bigr\vert\Omega \Bigr\rangle_0 ,\quad\label{-+} \\
i\Delta_{\scriptscriptstyle +-}(x;x^{\p}) \!\!\! &=& \!\!\!
\Bigl\langle \Omega \Bigl\vert \varphi(x^{\p}) \varphi(x) \Bigr\vert
\Omega \Bigr\rangle_0 , \quad \label{+-} \\
i\Delta_{\scriptscriptstyle --}(x;x^{\p}) \!\!& =& \!\!
\Bigl\langle \Omega \Bigl\vert \overline{T}\Bigl(\varphi(x)
\varphi(x^{\p}) \Bigr) \Bigr\vert \Omega \Bigr\rangle_0 . \label{--}
\end{eqnarray}
The subscript $0$ indicates vacuum expectation values in the free theory.
A careful reader might have noticed that the $++$ propagator is the usual
Feynman propagator and the $--$ one is its complex conjugate; the $-+$
propagator is similarly the conjugate of the $+-$ one.

We close by employing the Schwinger-Keldysh formalism to show
that a causal result is achieved in scalar field
theory with interaction $-\frac{1}{6}\lambda\varphi^3$.
To facilitate this simple computation we introduce the linearized
Schwinger-Keldysh effective field equation without
deriving it \cite{Miao,FW,TOW}\footnote{Although there are four
2-point 1PI (One particle irreducible) functions in the in-in formalism,
we only need two of them in the Schwinger-Keldysh effective equation.},
\begin{equation}
\frac{\delta \Gamma[\varphi_{\scriptscriptstyle
+},\varphi_{\scriptscriptstyle -}] }{\delta \varphi_{\scriptscriptstyle
+}(x)} \Biggl\vert_{\varphi_{\scriptscriptstyle \pm} = \varphi} \!\!\! =
\frac{\delta S[\varphi]}{\delta \varphi(x)} - \! \int \! d^4x^{\p}
\Bigl[M^2_{\scriptscriptstyle ++}\!(x;x^{\p}) +
M^2_{\scriptscriptstyle +-}\!(x;x^{\p})\Bigr] \varphi(x^{\p}).\label{SK}
\end{equation}
The two squared self-masses in $\varphi^3$ theory can be expressed as,
\begin{eqnarray}
M^2_{\scriptscriptstyle +\pm}(x;x')\!=\! \mp i\frac{\lambda^2}{2}
\Bigl[ i\Delta_{\scriptscriptstyle +\pm}(x;x') \Bigr]^2
\!=\!\mp i \frac{\lambda^2}{2}
\frac{\Gamma^2(\frac{D}{2}\!-\!1)}{16 \pi^{D}}
\Biggl[\frac{1}{\Delta x^2_{\scriptscriptstyle +\pm}(x;x')}
\Biggr]^{D-2},\label{TOT2pt}
\end{eqnarray}
and the two invariant intervals in the denominator of (\ref{TOT2pt}) are,
\begin{eqnarray}
&&\Delta x^2_{\scriptscriptstyle ++}(x;x')=\parallel\vec{x}-\vec{x}'\parallel^2
-(|t-t'|-i\delta)^2,\label{x++}\\
&&\Delta x^2_{\scriptscriptstyle +-}(x;x')=\parallel\vec{x}-\vec{x}'\parallel^2
-(t-t'+i\delta)^2.\label{x+-}
\end{eqnarray}

First of all, we notice that $\Delta x^2_{\scriptscriptstyle ++}$ equals
$\Delta x^2_{\scriptscriptstyle +-}$ while the time $t'$ is in the future of
the time $t$. A direct consequence of this is that the contribution from
$M^2_{\scriptscriptstyle ++}(x;x')$ cancels that from
$M^2_{\scriptscriptstyle +-}(x;x')$.
This implies no contributions from $t'$ in the future of the time $t$. Second,
when the time $t'$ lies in the past of the time $t$,
$\Delta x^2_{\scriptscriptstyle +-}(x;x')$ is the complex conjugate of
$\Delta x^2_{\scriptscriptstyle ++}(x;x')$, which indicates
$i\Delta_{\scriptscriptstyle +-}(x;x')=[i\Delta_{\scriptscriptstyle ++}(x;x')]^*$.
The combination of the two self-squared masses can be written as,
\begin{eqnarray}
\Bigl[M^2_{\scriptscriptstyle ++}\!+\!M^2_{\scriptscriptstyle +-}\Bigr](x;x')
\!=\!-i\frac{\lambda^2}{2} \Biggl\{\Bigl[i\Delta_{\scriptscriptstyle ++}(x;x')\Bigr]^2
\!\!-\!\Bigl([i\Delta_{\scriptscriptstyle ++}(x;x')]^*\Bigr)^2\Biggr\}
\!\!\longrightarrow \textrm{real}.\label{2M2}
\end{eqnarray}
One can infer from equation (\ref{2M2}) that all contributions from the past of
the time $t$ are real. Further, when the points $x^{\mu}$ and ${x'}^{\mu}$ are
spacelike separated the real parts of the invariant intervals are positive and the
different infinitesimal imaginary parts are irrelevant. Hence the $++$ and $+-$
contributions cancel.
In summary, we have established that the sum of $M^2_{\scriptscriptstyle ++}(x;x')$
and $M^2_{\scriptscriptstyle+-}(x;x')$ is zero except when ${x'}^{\mu}$ lies on or
within the past lightcone of $x^{\mu}$. Using the linearized Schwinger-Keldysh
effective equation (\ref{SK}) also guarantees that the result derived from it must be
real and causal.

\section{A worked-out example}
When one considers loop corrections to the scalar or tensor power spectra,
one inevitably needs higher order interaction vertices. Even though it is tedious
to obtain them from the gauge-fixed and constrained Lagrangian, several of them
have been worked out:
\begin{itemize}
\item{The $\zeta^3$ interaction by Maldacena \cite{JM};}
\item{Simple results for the $\zeta^4$ terms by Seery, Lidsey
and Sloth \cite{SLS};}
\item{The interactions of $\zeta^5$ and $\zeta^6$ discussed
by Jarnhus and Sloth \cite{JaSl};}
\item{The lowest $\zeta$--graviton interactions,
$\zeta h^2\;,\;\zeta^2 h$ and $\zeta^2 h^2$,
given by Xue, Gao and Brandenberger \cite{XGB}.}
\end{itemize}
Many diagrams are possible with these interactions but the simplest consists of
a single loop with two 3-point vertices. We lose nothing to consider a scalar
theory with a cubic interaction in flat spacetime,
\begin{eqnarray}
\mathcal{L}=-\frac{1}{2}\partial_{\mu}\varphi\partial_{\nu}\varphi g^{\mu\nu}
\!-\frac{\lambda}{3!}\varphi^3\;,\label{3scalar}
\end{eqnarray}
because the diagram topology is the same as for scalar-driven inflation but the
actual computation is vastly simpler.

In this section we use this worked-out example
to compute the one-loop correction to
the power spectrum. We employ both the mode function definition (\ref{modedef})
and the corrector definition (\ref{VEVdef}). What we show is that two definitions
disagree at one loop. The curious reader can find the explicit, and finite results
for each definition worked out in Appendix D.

\subsection{The mode function definition}
In this subsection we first give some identities to facilitate the computation.
We then use the linearized Schwinger-Keldysh effective field equation
to solve for the first order correction to the mode function. Finally the formal
expression for the corresponding power spectrum of $\varphi^3$ theory at one loop
is presented.

The correction to the power spectrum by definition (\ref{modedef})
at one loop order is,
\begin{eqnarray}
\Delta u_1(t,k)u^*(t,k)+ \Delta u^*_1(t,k)u(t,k)\;.\label{u1u0}
\end{eqnarray}
Here $u(t,k)$ is the tree order mode function.
Its relation with the free field expansion is,
\begin{eqnarray}
\varphi_0(t,\vec{x})=\!\!\int\!\!\frac{d^3k}{(2\pi)^3}\Biggl\{u(t,k)\alpha(k)
e^{i\vec{k}\cdot\vec{x}}+ u^*(t,k)\alpha^{\dagger}(k)
e^{-i\vec{k}\cdot\vec{x}}\Biggr\}\;.\label{field-0}
\end{eqnarray}
Applying (\ref{field-0}) to
(\ref{++}) - (\ref{--})
we obtain the propagators with different polarities in terms of the mode functions,
\begin{eqnarray}
&&i\Delta_{\scriptscriptstyle ++}(x;y)\!=
\!\!\int\!\!\frac{d^3k}{(2\pi)^3}e^{i\vec{k}\cdot(\vec{x}-\vec{y})}
\left\{\!\matrix{\theta(x^0\!\!-\!y^0)u(x^0,k)u^*(y^0,k)\cr
+\theta(y^0\!\!-\!x^0)u^*(x^0,k)u(y^0,k)}
\!\right\}\;,\label{++u} \\
&&i\Delta_{\scriptscriptstyle -+}(x;y)\!=
\!\!\int\!\!\frac{d^3k}{(2\pi)^3}e^{i\vec{k}\cdot(\vec{x}-\vec{y})}
u(x^0,k)u^*(y^0,k)\label{-+u}\;,\\
&&i\Delta_{\scriptscriptstyle +-}(x;y)\!=
\!\!\int\!\!\frac{d^3k}{(2\pi)^3}e^{-i\vec{k}\cdot(\vec{x}-\vec{y})}
u^*(x^0,k)u(y^0,k)=[i\Delta_{\scriptscriptstyle{-+}}(x;y)]^*
\;,\label{+-u}\\
&&i\Delta_{\scriptscriptstyle --}(x;y)=
[i\Delta_{\scriptscriptstyle{++}}(x;y)]^* \nonumber\\
&&\hspace{2cm}=\!\!\int\!\!\frac{d^3k}{(2\pi)^3}
e^{-i\vec{k}\cdot(\vec{x}-\vec{y})}
\left\{\!\matrix{\theta(x^0\!\!-\!y^0)u^*(x^0,k)u(y^0,k)\cr
+\theta(y^0\!\!-\!x^0)u(x^0,k)u^*(y^0,k)}
\!\right\}\;.\label{--u}
\end{eqnarray}
The symbol $\Delta u_1(t,k)$ in (\ref{u1u0}) denotes the first order correction to
the mode function. For convenience of later discussion we drop the
subscript of $\Delta u_1(t,k)$. It obeys,
\begin{eqnarray}
\mathcal{D}[\Delta u(t,k) e^{i\vec{k}\cdot\vec{x}}]-\!\int \!\!
d^4y [M^2_{\scriptscriptstyle ++}(x;y) +
M^2_{\scriptscriptstyle +-}(x;y)]
u(y^0,k)e^{i\vec{k}\cdot\vec{y}}=0\;,\label{u1eqn}
\end{eqnarray}
and can be solved formally,
\begin{eqnarray}
\Delta u(t,k)=\!\!\int\!\! d^4 y G_{\scriptscriptstyle{Ret}}(x;y)
\!\int\!\! d^4 y'[M^2_{\scriptscriptstyle ++}
\!+ M^2_{\scriptscriptstyle +-}](y;y')
e^{i\vec{k}\cdot(\vec{y'}-\vec{x})}u(y'^0,k)\;.\label{u1sol}
\end{eqnarray}
Here $G_{\scriptscriptstyle{Ret}}(x;y)$ is
the retarded Green's function for the operator
$\mathcal{D}$
and can be expressed in terms of the Schwinger-Keldysh propagators
(\ref{++}) - (\ref{--}),
\begin{eqnarray}
G_{\scriptscriptstyle{Ret}}(x;y)\!=\!-i[i\Delta_{\scriptscriptstyle{++}}
\!-i\Delta_{\scriptscriptstyle{+-}}](x;y)\;.\label{Gret}
\end{eqnarray}
Also note that the various $\pm$ polarities of the self-mass-squared for
$\varphi^3$ theory are,
\begin{eqnarray}
-iM^2_{\scriptscriptstyle \pm\pm}(y;y')\!=\!-\frac{\lambda^2}{2}
\Bigl[i\Delta_{\scriptscriptstyle \pm\pm}(y;y')\Bigr]^2 \;\;
,\;\;-iM^2_{\scriptscriptstyle \pm\mp}(y;y')\!=\!\frac{\lambda^2}{2}
\Bigl[i\Delta_{\scriptscriptstyle \pm\mp}(y;y')\Bigr]^2.\label{Msquare}
\end{eqnarray}
Inserting (\ref{u1sol}), (\ref{Gret}) and their complex conjugates given by
(\ref{+-u}), (\ref{--u}) to (\ref{u1u0}) we get,
\begin{eqnarray}
&&\hspace{-1.2cm}\Delta u(t,k) u^{*}(t,k) + \Delta u^{*}(t,k) u(t,k)=
\int\!\! d^4 y \!\int\!\! d^4 y'\times\nonumber\\
&&\hspace{-1.4cm}\left\{\!\!\matrix{[i\Delta_{\scriptscriptstyle ++}
\!-i\Delta_{\scriptscriptstyle +-}](x;y)
[-iM^2_{\scriptscriptstyle ++}
\!-iM^2_{\scriptscriptstyle +-}](y;y')
e^{i\vec{k}\cdot(\vec{y'}-\vec{x})}u^*(t,k)u(y'^0,k) \cr
+[i\Delta_{\scriptscriptstyle --}\!
-i\Delta_{\scriptscriptstyle -+}](x;y)
[-iM^2_{\scriptscriptstyle -+}
\!-iM^2_{\scriptscriptstyle --}](y;y')
e^{-i\vec{k}\cdot(\vec{y'}-\vec{x})}u(t,k)u^*(y'^0,k)}\!\!\right\}\!.
\label{Duu1}
\end{eqnarray}
Besides, there is no harm to shift the spatial coordinates in (\ref{Duu1}),
\begin{eqnarray}
\vec{y'}\longrightarrow \vec{y'} +\vec{x}\;\;\;;\;\;\;
\vec{y}\longrightarrow \vec{y} +\vec{x},
\end{eqnarray}
and it can be written as,
\begin{eqnarray}
&&\hspace{-1cm}\Delta u(t,k) u^{*}(t,k) + \Delta u^{*}(t,k) u(t,k)=
\int\!\! d^4 y \!\int\!\! d^4 y'e^{-i\vec{k}\cdot\vec{y'}}
\times\nonumber\\&&\hspace{-1cm}
\left\{\!\matrix{[i\Delta_{\scriptscriptstyle ++}
\!-i\Delta_{\scriptscriptstyle +-}](t,\vec{0};y)
[-iM^2_{\scriptscriptstyle ++}
\!-iM^2_{\scriptscriptstyle +-}](y;y')
u^*(t,k)u(y'^0,k) \cr
+[i\Delta_{\scriptscriptstyle --}\!
-i\Delta_{\scriptscriptstyle -+}](t,\vec{0};y)
[-iM^2_{\scriptscriptstyle -+}
\!-iM^2_{\scriptscriptstyle --}](y;y')
u(t,k)u^*(y'^0,k)}\!\!\right\}.\label{Duu2}
\end{eqnarray}

In the next step we employ the following identities\footnote{
$\Bigl<\varphi_0(x)\varphi_0(y)\Bigr>$ is the abbreviation of
$\Bigl<\Omega|\varphi_0(x)\varphi_0(y)|\Omega\Bigr>$.},
\begin{eqnarray}
&&\Bigl[i\Delta_{\scriptscriptstyle ++}\!-\!
i\Delta_{\scriptscriptstyle +-}\Bigr](x;y)\!=\!-
\Bigl[i\Delta_{\scriptscriptstyle --}\!-\!
i\Delta_{\scriptscriptstyle -+}\Bigr](x;y)\nonumber\\
&&\hspace{3cm}=\theta(x^0\!-\!y^0)\Bigl\{
\Bigl<\varphi_0(x)\varphi_0(y)\Bigr>
-\Bigl<\varphi_0(y)\varphi_0(x)\Bigr>\Bigr\},\\
&&[-iM^2_{\scriptscriptstyle ++}
\!-iM^2_{\scriptscriptstyle +-}](y;y')=
-[-iM^2_{\scriptscriptstyle -+}
\!-iM^2_{\scriptscriptstyle --}](y;y')\nonumber\\
&&\hspace{2cm}=-\frac{\lambda^2}{2}\theta(y^0-y'^0)
\Bigl\{\Bigl<\varphi_0(y)\varphi_0(y')\Bigr>^2
-\Bigl<\varphi_0(y')\varphi_0(y)\Bigr>^2\Bigr\},\label{M2Delta2}
\end{eqnarray}
in (\ref{Duu2}) and a further simplification is,
\begin{eqnarray}
&&\hspace{-1cm}\Delta u(t,k) u^{*}(t,k)
+ \Delta u^{*}(t,k) u(t,k)=\nonumber\\
&&\hspace{-1cm}-\frac{\lambda^2}{2}\!\!\int_0^{t}\!\!dy^0
\!\!\int_0^{y^0}\!\!\!dy'^0\!\!\int\!\! d^3 y \!\!\int\!\! d^3 y'
e^{-i\vec{k}\cdot\vec{y'}}
\Bigl[ \Bigl<\varphi_{0}(t,\vec{0})\varphi_0(y)\Bigr>
-\Bigl<\varphi_0(y)\varphi_0(t,\vec{0})\Bigr>\Bigr]\nonumber\\
&&\hspace{-1.2cm}\times\Bigl[\Bigl<\varphi_0(y)\varphi_0(y')\Bigr>^2
\!-\!\Bigl<\varphi_0(y')\varphi_0(y)\Bigr>^2\Bigr]
\Bigl[u(t,k)u^*(y'^0,k)+u^*(t,k)u(y'^0,k)\Bigr].\label{Duu3}
\end{eqnarray}

\subsection{The 2-point correlator definition}
In this subsection we compute the first order corrections to the power
spectrum by spatially Fourier transforming the 2-point correlators.
Within the in-in formalism the external legs of 2-point correlators could
have the following polarities: $(++), (-+), (+-)$ and $(--)$. We begin with
the $(-+)$ 2-point correlator and compute the power spectrum by employing
the correlator definition. We found that the result doesn't agree with
(\ref{Duu3}). We also show that none of the other in-in correlators, nor
any linear combination of them, can resolve the disagreement.

The spatial Fourier transform of the 2-point correlators
of $\varphi^3$ theory is,
\begin{eqnarray}
\int\!\! d^3 x e^{-i\vec{k}\cdot\vec{x}} \Bigl<\Omega|
\varphi(t,\vec{x})\varphi(t,\vec{0})|\Omega\Bigr>\;.
\end{eqnarray}
We begin with the $(-+)$ 2-point correlator at one loop order.
The generic diagram topology is depicted in Fig.~1. The explicit form is,
\begin{eqnarray}
&&\int d^3x e^{-i\vec{k}\cdot\vec{x}}
\int\!\! d^4 y \!\int\!\! d^4 y'
\left\{\!\matrix{+i\Delta_{\scriptscriptstyle -+}(x;y)
[-iM^2_{\scriptscriptstyle ++}(y;y')]
\,i\Delta_{\scriptscriptstyle ++}(x';y')\cr
+i\Delta_{\scriptscriptstyle -+}(x;y)
[-iM^2_{\scriptscriptstyle +-}(y;y')]
\,i\Delta_{\scriptscriptstyle +-}(x';y')\cr
+i\Delta_{\scriptscriptstyle --}(x;y)
[-iM^2_{\scriptscriptstyle -+}(y;y')]
\,i\Delta_{\scriptscriptstyle ++}(x';y')\cr
+i\Delta_{\scriptscriptstyle --}(x;y)
[-iM^2_{\scriptscriptstyle --}(y;y')]
\,i\Delta_{\scriptscriptstyle +-}(x';y')}\!\right\},\\
&&=\int\!\! d^4 y \!\int\!\! d^4 y'
e^{-i\vec{k}\cdot\vec{y}}\times\nonumber\\
&&\left\{\!\matrix{\hspace{-1.5cm}u(t,k)u^*(y^0,k)
\left\{\!\matrix{+[-iM^2_{\scriptscriptstyle ++}(y;y')]
\,i\Delta_{\scriptscriptstyle ++}(x';y')\cr
+[-iM^2_{\scriptscriptstyle +-}(y;y')]
\,i\Delta_{\scriptscriptstyle +-}(x';y')}\!\right\}\cr
+\Bigl\{\theta(t\!-\!y^0)u^*(t,k)u(y^0,k)
+\theta(y^0\!-\!t)u(t,k)u^*(y^0,k)\Bigr\}\cr
\hspace{4cm}\times\left\{\!\matrix{
+[-iM^2_{\scriptscriptstyle -+}(y;y')]
\,i\Delta_{\scriptscriptstyle ++}(x';y')\cr
+[-iM^2_{\scriptscriptstyle --}(y;y')]
\,i\Delta_{\scriptscriptstyle +-}(x';y')}\!\right\}}
\!\right\}.\label{4F1}
\end{eqnarray}
\begin{figure}
\begin{center}
\includegraphics[width=13.8cm]{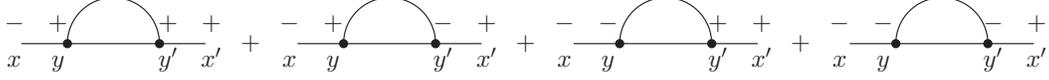}
\end{center}
\caption{One loop contribution to the $(-+)$ 2-point correlator.
We define the coordinates of the two external legs to be
$x^{\mu}=(t,\vec{x})$ and $x'^{\mu}=(t,\vec{0})$.}
\label{Figure1}
\end{figure}

After comparing (\ref{4F1}) with (\ref{Duu2}) we interchange $y$ with
$y'$ in (\ref{4F1}),
\begin{eqnarray}
&&\int\!\! d^4 y \!\int\!\! d^4 y'e^{-\vec{k}\cdot\vec{y'}}
\times\nonumber\\
&&\left\{\!\matrix{
\left\{\!\matrix{\,i\Delta_{\scriptscriptstyle ++}(t,\vec{0};y)
[-iM^2_{\scriptscriptstyle ++}(y;y')]\cr
\,+i\Delta_{\scriptscriptstyle +-}(t,\vec{0};y)
[-iM^2_{\scriptscriptstyle -+}(y;y')]}\!\right\}
u(t,k)u^*(y'^0,k)\cr
\hspace{-2.5cm}+\left\{\!\matrix{
\,i\Delta_{\scriptscriptstyle ++}(t,\vec{0};y)
[-iM^2_{\scriptscriptstyle +-}(y;y')]\cr
\,+i\Delta_{\scriptscriptstyle +-}(t,\vec{0};y)
[-iM^2_{\scriptscriptstyle --}(y;y')]}\!\right\}\times\cr
\Bigl\{\theta(t\!-\!y'^0)u^*(t,k)u(y'^0,k)
+\theta(y'^0\!-\!t)u(t,k)u^*(y'^0,k)\Bigr\}
}\!\right\}.\label{4F2}
\end{eqnarray}
Here we used $i\Delta_{\pm\mp}(y;x)=i\Delta_{\mp\pm}(x;y)$
and $i\Delta_{\pm\pm}(y;x)=i\Delta_{\pm\pm}(x;y)$.

To simplify (\ref{4F2}), we combine the first line with the third
and the second line with the fourth. We then extract out the common expression
from each of the combinations. The total result in (\ref{4F2}) consists of
two parts. One of them is proportional to $\theta(t-y^0)\theta(y^0-y'^0)$
and the other to $\theta(t-y'^0)\theta(y'^0-y^0)$. We could use these
theta functions to restrict the range of temporal integrations and
give the final expressions in a more concise form denoted by $(A)$ and $(B)$,
\begin{eqnarray}
&&(A)\!=\! -\frac{\lambda^2}{2}
\!\!\int_0^{t}\!dy^0\!\!\int_0^{y^0}\!\!dy'^0
\!\!\int\!\! d^3y\!\!\int\!\! d^3y'\!e^{-i\vec{k}\cdot\vec{y'}}
\Bigl[\Bigl<\varphi_0(t,\vec{0})\varphi_0(y)\Bigr>
\!-\Bigl<\varphi_0(y)\varphi_0(t,\vec{0})\Bigr>\Bigr]\nonumber\\
&&\times\Bigl[\Bigl<\varphi_0(y)\varphi_0(y')\Bigr>^2
u(t,k)u^*(y'^0,k)\!-\!\Bigl<\varphi_0(y')\varphi_0(y)\Bigr>^2
u^*(t,k)u(y'^0,k)\Bigr],\label{4Ftot1}\\
&&(B)\!=\!-\frac{\lambda^2}{2}
\!\!\int_0^t\!dy^0\!\!\int_{y^0}^t\!\!dy'^0
\!\!\int\!\! d^3y\!\!\int\!\! d^3y'\!e^{-i\vec{k}\cdot\vec{y'}}
\Bigl[u(t,k)u^*(y'^0\!,k)\!-\!u^*(t,k)u(y'^0\!,k)\Bigr]\nonumber\\
&&\times\Bigl[\Bigl<\varphi_0(y')\varphi_0(y)\Bigr>^2
\Bigl<\varphi_0(t,\vec{0})\varphi_0(y)\Bigr> \!-\!
\Bigl<\varphi_0(y)\varphi_0(y')\Bigr>^2
\Bigl<\varphi_0(y)\varphi_0(t,\vec{0})\Bigr>\Bigr].\label{4Ftot2}
\end{eqnarray}

In order to compare (\ref{4Ftot1}) $+$ (\ref{4Ftot2}) with (\ref{Duu3}),
we make several reformulations of (\ref{4Ftot1}) and (\ref{4Ftot2}).
At the first step we convert mode functions to the vacuum expectation value (VEV)
of the products of two fields. Equations (\ref{4Ftot1}) and (\ref{4Ftot2}) can be
written as,
\begin{eqnarray}
&&(A)\!=\! -\frac{\lambda^2}{2}
\!\!\int_0^{t}\!dy^0\!\!\int_0^{y^0}\!\!dy'^0
\!\!\int\!\! d^3y\!\!\int\!\! d^3y'\!\!\int\!\!d^3 x
e^{-i\vec{k}\cdot\vec{x}}\times\nonumber\\
&&\hspace{-1cm}\left\{\matrix{
\Bigl<\varphi_0(y)\varphi_0(y')\Bigr>^2
\Bigl<\varphi_0(t,\vec{x})\varphi_0(y')\Bigr>
\Bigl[\Bigl<\varphi_0(t,\vec{0})\varphi_0(y)\Bigr>
\!-\Bigl<\varphi_0(y)\varphi_0(t,\vec{0})\Bigr>\Bigr]\cr
\!\!+\Bigl<\varphi_0(y')\varphi_0(y)\Bigr>^2
\Bigl<\varphi_0(y')\varphi_0(t,\vec{x})\Bigr>
\Bigl[\Bigl<\varphi_0(y)\varphi_0(t,\vec{0})\Bigr>
\!-\!\Bigl<\varphi_0(t,\vec{0})\varphi_0(y)\Bigr>\Bigr]
}\!\!\right\}\!,
\label{4Ftot1a}\\
&&(B)\!=\!-\frac{\lambda^2}{2}
\!\!\int_0^t\!dy'^0\!\!\int_0^{y'^0}\!\!\!dy^0
\!\!\int\!\! d^3y\!\!\int\!\! d^3y'\!\!\int\!\!d^3 x
e^{-i\vec{k}\cdot\vec{x}}\times\nonumber\\
&&\hspace{-1.2cm}\left\{\matrix{
\Bigl<\varphi_0(y')\varphi_0(y)\Bigr>^2
\Bigl<\varphi_0(t,\vec{0})\varphi_0(y)\Bigr>
\Bigl[\Bigl<\varphi_0(t,\vec{x})\varphi_0(y')\Bigr>
\!-\Bigl<\varphi_0(y')\varphi_0(t,\vec{x})\Bigr>\Bigr]\cr
\!\!+\Bigl<\varphi_0(y)\varphi_0(y')\Bigr>^2
\Bigl<\varphi_0(y)\varphi_0(t,\vec{0})\Bigr>
\Bigl[\Bigl<\varphi_0(y')\varphi_0(t,\vec{x})\Bigr>
\!-\!\Bigl<\varphi_0(t,\vec{x})\varphi_0(y')\Bigr>\Bigr]
}\!\!\right\}\!.\label{4Ftot2a}
\end{eqnarray}
Note that we have rearranged the order of the temporal integrations
in (\ref{4Ftot2a}). Before executing the second step, we introduce
two key identities,
\begin{eqnarray}
&&\Bigl<\varphi(t,\vec{x})\varphi(y^0,\vec{y})\Bigr>=
\Bigl<\varphi(t,\vec{0})\varphi(y^0,\vec{y}\!-\!\vec{x})
\Bigr>\;,\label{id1} \\
&&\Bigl<\varphi(t,\vec{x})\varphi(y^0,\vec{0})\Bigr>=
\Bigl<\varphi(t,-\vec{x})\varphi(y^0,\vec{0})\Bigr>\;.\label{id2}
\end{eqnarray}
The identity (\ref{id1}) comes from spatial translation invariance and
the identity (\ref{id2}) is the consequence of spatial rotation invariance.

At the second stage we repeatedly apply (\ref{id1}) and (\ref{id2}) to the
contribution (B) in (\ref{4Ftot2a}) and leave (\ref{4Ftot1a}) unchanged.
The first manipulation we make is,
\begin{eqnarray}
&&\Bigl<\varphi_0(t,\vec{0})\varphi_0(y)\Bigr> =
\Bigl<\varphi_0(t,\vec{0})\varphi_0(y^0,-\vec{y})\Bigr>\;,\nonumber\\
&&\Bigl<\varphi_0(y)\varphi_0(t,\vec{0})\Bigr> =
\Bigl<\varphi_0(y^0,-\vec{y})\varphi_0(t,\vec{0})\Bigr>\;,
\end{eqnarray}
and then shift the spatial coordinates for all of the terms in (\ref{4Ftot2a}),
\begin{eqnarray}
\vec{y}\longrightarrow\vec{y}+\vec{x}\;\;,\;\;
\vec{y'}\longrightarrow\vec{y'}+\vec{x}.
\end{eqnarray}
This change would not affect the range of integrations or the VEV
of $\varphi_0(y)\varphi_0(y')$.  Here we only present what has been changed
by these transformations. The first part proportional to
$\!\Bigl<\!\varphi_0(y')\varphi_0(y)\!\Bigr>^2\!$ becomes,
\begin{eqnarray}
\hspace{-.5cm}\Bigl<\varphi_0(t,\vec{0})\varphi_0(y^0,-\vec{y}\!+\!\vec{x})\Bigr>
\Bigl[\Bigl<\varphi_0(t,\vec{x})\varphi_0(y'^0,\vec{y'}\!+\!\vec{x})\Bigr>
\!-\Bigl<\varphi_0(y'^0,\vec{y'}\!+\!\vec{x})
\varphi_0(t,\vec{x})\Bigr>\Bigr],\label{cha}
\end{eqnarray}
and the second part proportional to
$\!\Bigl<\!\varphi_0(y)\varphi_0(y')\!\Bigr>^2\!$
has been changed to,
\begin{eqnarray}
\Bigl<\varphi_0(y^0,-\vec{y}\!+\!\vec{x})\varphi_0(t,\vec{0})\Bigr>
\Bigl[\Bigl<\varphi_0(y'^0,\vec{y'}\!+\!\vec{x})\varphi_0(t,\vec{x})\Bigr>
\!-\!\Bigl<\varphi_0(t,\vec{x})\varphi_0(y'^0,\vec{y'}\!+\!\vec{x})
\Bigr>\Bigr].\label{chb}
\end{eqnarray}

In the next step we apply first (\ref{id2}) and then (\ref{id1})
to the first term of (\ref{cha}) and (\ref{chb}),
\begin{eqnarray}
&&\Bigl<\varphi_0(t,\vec{0})\varphi_0(y^0,-\vec{y}\!+\!\vec{x})\Bigr>
\!=\!\Bigl<\varphi_0(t,\vec{0})\varphi_0(y^0,\vec{y}\!-\!\vec{x})\Bigr>
\!=\!\Bigl<\varphi_0(t,\vec{x})\varphi_0(y^0,\vec{y})\Bigr>,\nonumber\\
&&\hspace{-.5cm}\Bigl<\varphi_0(y^0,-\vec{y}
\!+\!\vec{x})\varphi_0(t,\vec{0})\Bigr>
\!=\!\Bigl<\varphi_0(y^0,\vec{y}\!-\!\vec{x})\varphi_0(t,\vec{0})\Bigr>
\!=\!\Bigl<\varphi_0(y^0,\vec{y})\varphi_0(t,\vec{x})\Bigr>,
\end{eqnarray}
and employ spatial translation invariance (\ref{id1}) in the remaining terms of
(\ref{cha}) and (\ref{chb}). Take the final two terms of (\ref{cha}) as an
example,
\begin{eqnarray}
&&\Bigl[\Bigl<\varphi_0(t,\vec{x})\varphi_0(y'^0,\vec{y'}\!+\!\vec{x})\Bigr>
\!-\Bigl<\varphi_0(y'^0,\vec{y'}\!+\!\vec{x})
\varphi_0(t,\vec{x})\Bigr>\Bigr]\nonumber\\
&&\longrightarrow\Bigl[\Bigl<\varphi_0(t,\vec{0})
\varphi_0(y'^0,\vec{y'})\Bigr>
\!-\Bigl<\varphi_0(y'^0,\vec{y'})\varphi_0(t,\vec{0})\Bigr>\Bigr].
\end{eqnarray}
After gathering all manipulations we made so far,
the contribution (\ref{4Ftot2a}) can be expressed as,
\begin{eqnarray}
&&(B)\!=\!-\frac{\lambda^2}{2}
\!\!\int_0^t\!dy'^0\!\!\int_0^{y'^0}\!\!\!dy^0
\!\!\int\!\! d^3y\!\!\int\!\! d^3y'\!\!\int\!\!d^3 x
e^{-i\vec{k}\cdot\vec{x}}\times\nonumber\\
&&\hspace{-1cm}\left\{\matrix{
\Bigl<\varphi_0(y')\varphi_0(y)\Bigr>^2
\Bigl<\varphi_0(t,\vec{x})\varphi_0(y)\Bigr>
\Bigl[\Bigl<\varphi_0(t,\vec{0})\varphi_0(y')\Bigr>
\!-\Bigl<\varphi_0(y')\varphi_0(t,\vec{0})\Bigr>\Bigr]\cr
\!\!+\Bigl<\varphi_0(y)\varphi_0(y')\Bigr>^2
\Bigl<\varphi_0(y)\varphi_0(t,\vec{x})\Bigr>
\Bigl[\Bigl<\varphi_0(y')\varphi_0(t,\vec{0})\Bigr>
\!-\!\Bigl<\varphi_0(t,\vec{0})\varphi_0(y')\Bigr>\Bigr]
}\!\!\right\}\!.\label{4Ftot2b}
\end{eqnarray}
At the final step we interchange $y$ with $y'$ in (\ref{4Ftot2b}). It
turns out that the outcome is exactly the same as the one in (\ref{4Ftot1a}).
This means that the contribution $(A)$ precisely equals $(B)$.
Hence the total result could be written as $2\times(\ref{4Ftot1})$ or
$2\times(\ref{4Ftot1a})$. We choose the form which is close to
the expression derived from the mode function definition (\ref{Duu3}),
\begin{eqnarray}
&&(A)\!+\!(B)\!=\! -\frac{\lambda^2}{2}
\!\!\int_0^{t}\!dy^0\!\!\int_0^{y^0}\!\!dy'^0
\!\!\int\!\! d^3y\!\!\int\!\! d^3y'\!e^{-i\vec{k}\cdot\vec{y'}}
\Bigl[\Bigl<\varphi_0(t,\vec{0})\varphi_0(y)\Bigr>
\!-\Bigl<\varphi_0(y)\varphi_0(t,\vec{0})\Bigr>\Bigr]\nonumber\\
&&\times\Bigl[2\Bigl<\varphi_0(y)\varphi_0(y')\Bigr>^2
u(t,k)u^*(y'^0,k)\!-\!2\Bigl<\varphi_0(y')\varphi_0(y)\Bigr>^2
u^*(t,k)u(y'^0,k)\Bigr].\label{4FTOT}
\end{eqnarray}
Equations (\ref{4FTOT}) and (\ref{Duu3}) both have the same
integrations and the common factor
 $\Bigl[\Bigl<\varphi_0(t,\vec{0})\varphi_0(y)\Bigr>
\!-\Bigl<\varphi_0(y)\varphi_0(t,\vec{0})\Bigr>\Bigr]$
so we could just focus on the rest of the integrands.
The two integrands differ by having the factors,
\begin{eqnarray}
\Bigl<\varphi_0(y)\varphi_0(y')\Bigr>^2u^*(t,k)u(y'^0,k)\;\textrm{and}\;
-\Bigl<\varphi_0(y')\varphi_0(y)\Bigr>^2u(t,k)u^*(y'^0,k),
\end{eqnarray}
in equation (\ref{Duu3}) replaced with
\begin{eqnarray}
\Bigl<\varphi_0(y)\varphi_0(y')\Bigr>^2 u(t,k)u^*(y'^0,k)\;\textrm{and}\;
-\Bigl<\varphi_0(y')\varphi_0(y)\Bigr>^2u^*(t,k)u(y'^0,k)\;.
\end{eqnarray}
Therefore we conclude that the mode function definition disagrees with
the spatial Fourier transform of the $(-+)$ 2-point correlator
at one loop.

We close this subsection by exploring the other Schwinger-Keldysh
correlators. First of all, we summarize several key points
learned from the reduction of the $(-+)$ 2-point correlator,
\begin{itemize}
\item{The contribution (A) in (\ref{4Ftot1}) equals (B) in (\ref{4Ftot2})
implies,
\begin{eqnarray}
\lefteqn{\int\!\!d^4y\!\!\int\!\!d^4y'\Bigl[
\theta(t\!-\!y^0)\theta(y^0\!-\!y'^0)\!+\!
\theta(t\!-\!y'^0)\theta(y'^0\!-\!y^0)\Bigr]}\nonumber\\
&\longrightarrow & 2\!\!\int_0^{t}\!dy^0\!\!\int_0^{y^0}\!\!\!dy'^0
\!\!\int\!\! d^3y\!\!\int\!\! d^3y'\!=2\!\int\!\!d^4y\!\!
\int\!\!d^4y'\theta(t\!-\!y^0)\theta(y^0\!-\!y'^0).
\end{eqnarray}}
\item{The same theta functions, $\theta(t\!-\!y^0)\theta(y^0\!-\!y'^0)$
and $\theta(t\!-\!y'^0)\theta(y'^0\!-\!y^0)$,
appear as well in the $(+-)$, $(++)$ and $(--)$ correlators.}
\item{Based on these two facts, we lose nothing by imposing the
time ordering $t>y^0>y'^0$ before making any further simplification.}
\item{One can further infer,
\begin{eqnarray}
&&\theta(t\!-\!y^0)\theta(y^0\!-\!y'^0)\Bigl\{i\Delta_{\scriptscriptstyle ++}(x;y)
\!=\!i\Delta_{\scriptscriptstyle -+}(x;y)\Bigr\}\;,\label{++-+}\\
&&\theta(t\!-\!y^0)\theta(y^0\!-\!y'^0)\Bigl\{i\Delta_{\scriptscriptstyle ++}(x';y')
\!=\!i\Delta_{\scriptscriptstyle -+}(x';y')\Bigr\}\;,\label{++-+pr}\\
&&\theta(t\!-\!y^0)\theta(y^0\!-\!y'^0)\Bigl\{i\Delta_{\scriptscriptstyle --}(x;y)
\!=\!i\Delta_{\scriptscriptstyle +-}(x;y)\Bigr\}\;,\label{--+-}\\
&&\theta(t\!-\!y^0)\theta(y^0\!-\!y'^0)\Bigl\{i\Delta_{\scriptscriptstyle --}(x';y')
\!=\!i\Delta_{\scriptscriptstyle +-}(x';y')\Bigr\}\;.\label{--+-pr}
\end{eqnarray}}
\end{itemize}

Second, we employ the rules developed in the preceding paragraph.
It is convenient to display all of the distinct forms for the spatial
Fourier transform of the in-in correctors. Because each of these forms
has the same integration
$2\!\!\int\!d^3x e^{-i\vec{k}\cdot\vec{x}}\!\!\int\!\! d^4 y \!\int\!\! d^4 y'$,
it is enough to only list the integrand,\\
from the $(-+)$ 2-point correlator,
\begin{eqnarray}
\theta(t\!-\!y^0)\theta(y^0\!-\!y'^0)
\left\{\!\matrix{+i\Delta_{\scriptscriptstyle -+}(x;y)
[-iM^2_{\scriptscriptstyle ++}(y;y')]
\,i\Delta_{\scriptscriptstyle ++}(x';y')\cr
+i\Delta_{\scriptscriptstyle -+}(x;y)
[-iM^2_{\scriptscriptstyle +-}(y;y')]
\,i\Delta_{\scriptscriptstyle +-}(x';y')\cr
+i\Delta_{\scriptscriptstyle --}(x;y)
[-iM^2_{\scriptscriptstyle -+}(y;y')]
\,i\Delta_{\scriptscriptstyle ++}(x';y')\cr
+i\Delta_{\scriptscriptstyle --}(x;y)
[-iM^2_{\scriptscriptstyle --}(y;y')]
\,i\Delta_{\scriptscriptstyle +-}(x';y')}\!\right\};\label{4F-+}
\end{eqnarray}
from the $(+-)$ 2-point correlator,
\begin{eqnarray}
\theta(t\!-\!y^0)\theta(y^0\!-\!y'^0)
\left\{\!\matrix{+i\Delta_{\scriptscriptstyle ++}(x;y)
[-iM^2_{\scriptscriptstyle ++}(y;y')]
\,i\Delta_{\scriptscriptstyle -+}(x';y')\cr
+i\Delta_{\scriptscriptstyle ++}(x;y)
[-iM^2_{\scriptscriptstyle +-}(y;y')]
\,i\Delta_{\scriptscriptstyle --}(x';y')\cr
+i\Delta_{\scriptscriptstyle +-}(x;y)
[-iM^2_{\scriptscriptstyle -+}(y;y')]
\,i\Delta_{\scriptscriptstyle -+}(x';y')\cr
+i\Delta_{\scriptscriptstyle +-}(x;y)
[-iM^2_{\scriptscriptstyle --}(y;y')]
\,i\Delta_{\scriptscriptstyle --}(x';y')}\!\right\};\label{4F+-}
\end{eqnarray}
from the $(++)$ 2-point correlator,
\begin{eqnarray}
\theta(t\!-\!y^0)\theta(y^0\!-\!y'^0)
\left\{\!\matrix{+i\Delta_{\scriptscriptstyle ++}(x;y)
[-iM^2_{\scriptscriptstyle ++}(y;y')]
\,i\Delta_{\scriptscriptstyle ++}(x';y')\cr
+i\Delta_{\scriptscriptstyle ++}(x;y)
[-iM^2_{\scriptscriptstyle +-}(y;y')]
\,i\Delta_{\scriptscriptstyle +-}(x';y')\cr
+i\Delta_{\scriptscriptstyle +-}(x;y)
[-iM^2_{\scriptscriptstyle -+}(y;y')]
\,i\Delta_{\scriptscriptstyle ++}(x';y')\cr
+i\Delta_{\scriptscriptstyle +-}(x;y)
[-iM^2_{\scriptscriptstyle --}(y;y')]
\,i\Delta_{\scriptscriptstyle +-}(x';y')}\!\right\};\label{4F++}
\end{eqnarray}
from the $(--)$ 2-point correlator,
\begin{eqnarray}
\theta(t\!-\!y^0)\theta(y^0\!-\!y'^0)
\left\{\!\matrix{+i\Delta_{\scriptscriptstyle -+}(x;y)
[-iM^2_{\scriptscriptstyle ++}(y;y')]
\,i\Delta_{\scriptscriptstyle -+}(x';y')\cr
+i\Delta_{\scriptscriptstyle -+}(x;y)
[-iM^2_{\scriptscriptstyle +-}(y;y')]
\,i\Delta_{\scriptscriptstyle --}(x';y')\cr
+i\Delta_{\scriptscriptstyle --}(x;y)
[-iM^2_{\scriptscriptstyle -+}(y;y')]
\,i\Delta_{\scriptscriptstyle -+}(x';y')\cr
+i\Delta_{\scriptscriptstyle --}(x;y)
[-iM^2_{\scriptscriptstyle --}(y;y')]
\,i\Delta_{\scriptscriptstyle --}(x';y')}\!\right\}.\label{4F--}
\end{eqnarray}
Applying the relations (\ref{++-+}), (\ref{++-+pr}) and (\ref{--+-pr}) to the first
two lines of (\ref{4F+-}) and the relations (\ref{++-+pr}), (\ref{--+-}) and (\ref{--+-pr})
to the bottom two lines of (\ref{4F+-}), the integrands of the $(+-)$ and $(-+)$
2-point correlators reach the same form.
The differences between (\ref{4F++}) and (\ref{4F-+}) are those 2-point
functions propagating between $x$ and $y$. They become identical after the relations
(\ref{++-+}) and (\ref{--+-}) are employed. Expression (\ref{4F--}) also differs from
(\ref{4F+-}) by the propagators between the coordinates $x$ and $y$, and they
agree with each other after the same reduction as in the previous case is employed.

What we have just observed implies that the integrands of the four Schwinger-Keldysh
correlators reach the same expression after imposing the time ordering $t>y^0>y'^0$.
An alert reader might also have noticed that enforcing relations
(\ref{++-+})-(\ref{--+-pr}) directly to each term of equations
(\ref{4F-+})-(\ref{4F--}) all gives,
\begin{eqnarray}
\theta(t\!-\!y^0)\theta(y^0\!-\!y'^0)
\left\{\!\matrix{+i\Delta_{\scriptscriptstyle -+}(x;y)
[-iM^2_{\scriptscriptstyle ++}(y;y')]
\,i\Delta_{\scriptscriptstyle -+}(x';y')\cr
+i\Delta_{\scriptscriptstyle -+}(x;y)
[-iM^2_{\scriptscriptstyle +-}(y;y')]
\,i\Delta_{\scriptscriptstyle +-}(x';y')\cr
+i\Delta_{\scriptscriptstyle +-}(x;y)
[-iM^2_{\scriptscriptstyle -+}(y;y')]
\,i\Delta_{\scriptscriptstyle -+}(x';y')\cr
+i\Delta_{\scriptscriptstyle +-}(x;y)
[-iM^2_{\scriptscriptstyle --}(y;y')]
\,i\Delta_{\scriptscriptstyle +-}(x';y')}\!\right\}.\label{4FSame}
\end{eqnarray}
Recall that equations (\ref{4F-+})-(\ref{4F--}) have the same integration,
\begin{eqnarray*}
2\!\!\int\!d^3x e^{-i\vec{k}\cdot\vec{x}}\!\!\int\!\! d^4 y \!\int\!\! d^4 y' \;.
\end{eqnarray*}
Hence spatially Fourier transforming all the 1-loop Schwinger-Keldysh correlators
gives the same answer displayed in (\ref{4FTOT}). Even making a linear combination
of them would not compensate for all the terms in (\ref{Duu3}).
Therefore we have explicitly demonstrated that the 2-point correlator definition
disagrees with the mode function definition at one loop.

One can also obtain a simple form for the difference between the
one loop correction to the 2-point correlator (\ref{4FTOT}) and
the one loop correction to the definition based on the mode function (\ref{Duu3}),
\begin{eqnarray}
&&-\frac{\lambda^2}{2}\!\!\int\!\!d^3x e^{-i\vec{k}\cdot\vec{x}}
\!\!\int\!\! d^4y\!\!\int\!\! d^4y'\!
\theta(t\!-\!y^0)\theta(y^0\!-\!y'^0)
\Bigl[\Bigl<\varphi_0(y)\varphi_0(y')\Bigr>^2\!+\!
\Bigl<\varphi_0(y')\varphi_0(y)\Bigr>^2\Bigr]\nonumber\\
&&\hspace{-.3cm}\Bigl[\Bigl<\varphi_0(t,\vec{0})\varphi_0(y)\Bigr>\!-\!
\Bigl<\varphi_0(y)\varphi_0(t,\vec{0})\Bigr>\Bigr]
\Bigl[\Bigl<\varphi_0(t,\vec{x})\varphi_0(y')\Bigr>\!-\!
\Bigl<\varphi_0(y')\varphi_0(t,\vec{x})\Bigr>\Bigr].\label{diff}
\end{eqnarray}
Also note that this difference involves two retarded Green's
functions\footnote{ (\ref{diff}) seems to have a similar structure as
equation (4.39) in \cite{TensorDiv} and (C18) in \cite{HRV} for the in-in correlator of
the metric perturbations including loop corrections from matter fields. It corresponds to
the contribution named as ``induced fluctuations''.},
\begin{eqnarray}
&&-i\theta(t\!-\!y^0)\Bigl[\Bigl<\varphi_0(t,\vec{0})\varphi_0(y)\Bigr>
\!-\!\Bigl<\varphi_0(y)\varphi_0(t,\vec{0})\Bigr>\Bigr]\;\nonumber\\
&&-i\theta(t\!-\!y'^0)\Bigl[\Bigl<\varphi_0(t,\vec{x})\varphi_0(y')\Bigr>\!-\!
\Bigl<\varphi_0(y')\varphi_0(t,\vec{x})\Bigr>\Bigr].\label{2Gret}
\end{eqnarray}

\section{Epilogue}

We have analyzed loop corrections to two different ways of defining the primordial
power spectra which happen to coincide at tree order (\ref{treeid}). One of these
definitions involves the norm squared of the mode functions (\ref{modedef}). It can
be generalized by using the linearized Schwinger-Keldysh effective field equation
to quantum correct the mode function. The other definition involves the spatial
Fourier transform of the 2-point correlator (\ref{VEVdef}), which is generalized by
simply computing the correlator to higher orders using the Schwinger-Keldysh formalism.
To simplify our analysis we employed $\varphi^3$ theory in flat space.
The fact that its interactions have the same topology as those of inflationary cosmology
justifies this simplification. What we have found is that
the two definitions do not agree even at one loop order.

The 2-point correlator definition has the advantage of representing each power
spectrum in terms of a single expectation value. However, it has been claimed that
the coincident limit of its spatial Fourier transform is singular in the gravitational
case \cite{TensorDiv}. That implies that this quantity suffers from a new sort of
ultraviolet divergence, beyond the usual ones which BPHZ renormalization
absorbs. No one currently understands how to remove this new divergence;
at a minimum it would require a composite operator renormalization. This means
that the tensor power spectrum, for example, cannot be based on the
expectation value of $h_{ij}(t,\vec{x}) h_{ij}(t,\vec{0})$ but rather this plus
some higher order operator with which its one loop corrections mix. Even more disturbing,
from the perspective of cosmology, this new divergence arises
from LATE TIME correlations between fluctuations of matter fields, rather than
from anything that happened during primordial inflation. We believe
these late time effects should be removed, the same way one edits out
infrared radiation from Jupiter, the galactic plane, and other known sources,
and the same way the observed spectrum --- with its acoustic oscillations ---
is fitted using the late time transfer function to infer the almost
perfectly scale invariant primordial spectrum.

It is therefore reasonable to pursue alternatives to the usual definition
of the power spectrum. The mode function definition, which can not
be expressed as a single VEV, does seem strange at first, but a closer look
reveals some advantages. First, it is free of the late time artifacts once
the appropriate 1PI 2-point function has been renormalized.
Second, it is arguably a reasonable translation of what we should be doing.
CMB photons do not acquire their redshifts at the surface of last scattering
but rather by propagating through the perturbed geometry between the surface
of last scattering and the late time observer. The original computation by Sachs
and Wolfe \cite{SaWo} expresses the temperature fluctuation as an integral of
the metric perturbations along the photon's worldline. The time evolution for
these metric perturbations comes from solving the linearized equations in
the background geometry of late times, but the initial conditions come from
primordial inflation. It seems at least as reasonable to take these
initial conditions from the quantum corrected mode functions as
from the correlator.

What we are interested in is what theoretical objects represent the
``primordial power spectra''. How we define the power spectra is still
an open question. Loop corrections to them might be observable in the far future,
assuming that theorists can find a unique model of inflation to fix
the tree order prediction, that astronomers can measure the matter power spectra
in 3 dimensions, and that astrophysicists can develop expertise needed to extract the
primordial signal from foregrounds.

\centerline{\bf Acknowledgements}

We have profited from conversations on this subject with M. Fr\"{o}b,
A. Roura and R. P. Woodard. S.P.M. was supported by NWO Veni Project
\# 680-47-406. S.P. is supported by the Eberly Research Funds of The
Pennsylvania State University. The Institute for Gravitation and the Cosmos
is supported by the Eberly College of Science and the Office of the Senior
Vice President for Research at the Pennsylvania State University.
S.P.M. is grateful for the hospitality of the University of Crete
where the final draft was prepared.
S.P. acknowledges the hospitality of the University of Utrecht
where the main computation was conducted.

\begin{appendices}

\section{The canonical relation for the mode function}\label{cano-mode}

In this section we elucidate the relation between the Heisenberg operator and
its mode function using the cubic interaction in our worked-out example. We start with
solving for the field operator perturbatively from the equation of motion and then
compute the expectation value of its commutator with the creation operator.
The section closes by giving the canonical relation for the mode function
which is solved from the Schwinger-Keldysh effective field equation.

The equation of motion for the case we consider (\ref{3scalar}) is,
\begin{eqnarray}
\partial^2\varphi=\frac{\lambda}{2}\varphi^2.
\end{eqnarray}
Solving for the field operator perturbatively means,
\begin{eqnarray}
&&\varphi=\varphi_{\scriptscriptstyle 0}+\lambda\varphi_{\scriptscriptstyle 1}
+\lambda^2\varphi_{\scriptscriptstyle 2}+O(\lambda^3),
\end{eqnarray}
and the field operators at the order of $\lambda^0$, $\lambda^1$ and $\lambda^2$ obey,
\begin{eqnarray}
\partial^2\varphi_{\scriptscriptstyle 0}=0\;\;,\;\;
\partial^2\varphi_{\scriptscriptstyle 1}=
\frac{1}{2}\varphi_{\scriptscriptstyle 0}^{2}\;\;,\;\;
\partial^2\varphi_{\scriptscriptstyle 2}=\frac{1}{2}
\Bigl[\varphi_{\scriptscriptstyle 0}\varphi_{\scriptscriptstyle 1}
\!+\!\varphi_{\scriptscriptstyle 1}\varphi_{\scriptscriptstyle 0}\Bigr].
\end{eqnarray}
Note that we consider the initial value data to be zeroth order so that
$\varphi_{\scriptscriptstyle 1}$ and $\varphi_{\scriptscriptstyle 2}$ etc.
all vanish at $t=0$, as do their first time derivatives.
Hence $\varphi_{\scriptscriptstyle 1},\varphi_{\scriptscriptstyle 2}$
can be expressed in terms of $\varphi_{\scriptscriptstyle 0}$,
\begin{eqnarray}
&&\varphi_{\scriptscriptstyle 1}(x)\!=\!\frac12\!\int\!\!d^4y G_{\scriptscriptstyle Ret}(x;y)
\varphi_{\scriptscriptstyle 0}^2(y),\nonumber\\
&&\hspace{-.4cm}\varphi_{\scriptscriptstyle 2}(x)\!=\!\frac14\!\int\!\!d^4y\!\!\int\!\!d^4y'
G_{\scriptscriptstyle Ret}(x;y)G_{\scriptscriptstyle Ret}(y;y')
\Bigl[\varphi_{\scriptscriptstyle 0}(y)\varphi^2_{\scriptscriptstyle 0}(y')
\!+\!\varphi^2_{\scriptscriptstyle 0}(y')\varphi_{\scriptscriptstyle 0}(y)\Bigr].
\end{eqnarray}
Even though this theory is not free, we can still organize
the initial values of the full field and its first time derivative
in terms of free creation and annihilation operators,
\begin{eqnarray}
\alpha(\vec{k})\!\equiv\!\frac{\dot{u}^*(0)\widetilde{\varphi}(0,\vec{k})
\!-\!u^*(0)\dot{\widetilde{\varphi}}(0,\vec{k})}{u(0)\dot{u}^*(0)-\dot{u}(0)u^*(0)}
\;;\;\alpha^{\dagger}(\vec{k})\!\equiv\!
\Biggl[\frac{\dot{u}^*(0)\widetilde{\varphi}(0,\vec{k})
\!-\!u^*(0)\dot{\widetilde{\varphi}}(0,\vec{k})}
{u(0)\dot{u}^*(0)-\dot{u}(0)u^*(0)}\Biggr]^{\dagger}\!\!.\;
\end{eqnarray}
These operators define the free ground state $\vert \Omega \rangle$,
\begin{eqnarray}
\alpha|\Omega\Bigr>=0=\Bigl<\Omega|\alpha^{\dagger}\;\;\;\;,\;\;\;\;
\Bigl<\Omega|\Omega\Bigr>=1\,.
\end{eqnarray}
The mode function is the matrix element of the full field between
the $t=0$ free vacuum and the $t=0$ free one particle state,
$\langle \Omega \vert \varphi \alpha^{\dagger} \vert \Omega \rangle$.
We first commute the $\alpha^{\dagger}$ through the $\varphi$,
\begin{eqnarray}
&&\Bigl[\varphi(t,\vec{x}), \alpha^{\dagger}\Bigr]
=\Bigl[\varphi_{\scriptscriptstyle 0}(t,\vec{x})
\!+\!\lambda\varphi_{\scriptscriptstyle 1}(t,\vec{x})\!+\!\lambda^2
\varphi_{\scriptscriptstyle 2}(t,\vec{x})
\!+\!O(\lambda^3), \alpha^{\dagger}\Bigr]\nonumber\\
&&=\Phi_{\scriptscriptstyle 0}(t,\vec{x})\!+\!\lambda\!\!\int\!\!d^4y
G_{\scriptscriptstyle Ret}(x;y)\varphi_{\scriptscriptstyle 0}(y)\Phi_{\scriptscriptstyle 0}(y)
\!+\!\frac{\lambda^2}{2}\!\!\int\!\!d^4y\!\!\int\!\!d^4y'G_{\scriptscriptstyle Ret}(x;y)
G_{\scriptscriptstyle Ret}(y;y')\nonumber\\
&&\hspace{3cm}\times\Biggl\{\varphi_{\scriptscriptstyle 0}^2(y')\Phi_{\scriptscriptstyle 0}(y)\!+\!
\Bigl[\varphi_{\scriptscriptstyle 0}(y)\varphi_{\scriptscriptstyle 0}(y')
\!+\!\varphi_{\scriptscriptstyle 0}(y')\varphi_{\scriptscriptstyle 0}(y)\Bigr]
\Phi_{\scriptscriptstyle 0}(y')\Biggr\},\label{commutator}
\end{eqnarray}
where $\Phi_{\scriptscriptstyle 0}(t,\vec{x})$ is a c-number,
\begin{eqnarray}
\Phi_{\scriptscriptstyle 0}(t,\vec{x})
=\Bigl[\varphi_{\scriptscriptstyle 0}(t,\vec{x}), \alpha^{\dagger}\Bigr]
=u(t,k)e^{i\vec{k}\cdot\vec{x}}.\label{Phi0}
\end{eqnarray}
Taking the expectation value of equation (\ref{commutator}) actually simplifies
the expression because the second term with a single integral vanishes and
the first term of the final line is properly by a vacuum shift. For our purpose
only the last two terms matter,
\begin{eqnarray}
&&\hspace{-1cm}\Bigl<\Omega|\Bigl[\lambda^2\varphi_{\scriptscriptstyle 2},
\alpha^{\dagger}\Bigr]|\Omega\Bigr>\!=\!\frac{\lambda^2}{2}\!\!\int\!\!d^4y\!\!\int\!\!d^4y'
G_{\scriptscriptstyle Ret}(x;y)G_{\scriptscriptstyle Ret}(y;y')
\Bigl[i\Delta_{\scriptscriptstyle -+}\!+\!i\Delta_{\scriptscriptstyle +-}
\Bigr](y;y')\Phi_{\scriptscriptstyle 0}(y')\nonumber\\
&&\hspace{.8cm}=\!-i\frac{\lambda^2}{2}\!\!\int\!\!d^4y\!\!\int\!\!d^4y'\theta(y^0\!\!-\!y'^0)
G_{\scriptscriptstyle Ret}(x;y)\Bigl[i\Delta^2_{\scriptscriptstyle -+}\!-\!
i\Delta^2_{\scriptscriptstyle +-}\Bigr](y;y')\Phi_{\scriptscriptstyle 0}(y')\,.
\label{cano-op}
\end{eqnarray}
Here the second equality is obtained by employing equation (\ref{Gret}).

Our quantum corrected mode function comes from solving the linearized effective field
equation (\ref{u1eqn}). For convenience, we re-write equation (\ref{u1sol}) as,
\begin{eqnarray}
\Delta u(t,k)e^{i\vec{k}\cdot\vec{x}}=\!\!\int\!\! d^4 y
G_{\scriptscriptstyle{Ret}}(x;y)\!\int\!\! d^4 y'[M^2_{\scriptscriptstyle ++}
\!+ M^2_{\scriptscriptstyle +-}](y;y')u(y'^0,k)e^{i\vec{k}\cdot\vec{y'}}\;,
\label{u1sol1a}
\end{eqnarray}
and applying (\ref{M2Delta2}) and (\ref{Phi0}) to (\ref{u1sol1a}) gives,
\begin{eqnarray}
\hspace{-.5cm}\Delta u(t,k)e^{i\vec{k}\cdot\vec{x}}\!=\!-i\frac{\lambda^2}{2}
\!\!\int\!\! d^4 y\!\int\!\! d^4 y'\theta(y^0\!\!-\!y'^0)G_{\scriptscriptstyle{Ret}}(x;y)
\Bigl[i\Delta^2_{\scriptscriptstyle -+}
\!-\! i\Delta^2_{\scriptscriptstyle +-}\Bigr](y;y')
\Phi_{\scriptscriptstyle 0}(y').\label{modeexact}
\end{eqnarray}
Comparing equation (\ref{cano-op}) with (\ref{modeexact}) demonstrates the
canonical relation for the quantum corrected mode function,
\begin{eqnarray}
\Delta u(t,k)e^{i\vec{k}\cdot\vec{x}}\!=\!
\Bigl<\Omega|\Bigl[\lambda^2\varphi_{\scriptscriptstyle 2}(t,\vec{x}),
\alpha^{\dagger}\Bigr]|\Omega\Bigr>\;.\label{modeOP}
\end{eqnarray}

\section{The diagram topology for quantum corrections to the mode function}\label{topology}

\begin{figure}
\begin{center}
\includegraphics[width=4cm]{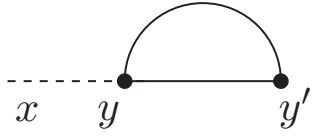}
\end{center}
\caption{This diagram characterizes the partial topology of the mode function (\ref{u1sol1a}).
The external leg denoted by a dashed line comes from the retarded Green's function whereas
the remaining part is the 2-point 1PI diagram.}
\label{Figure2}
\end{figure}

In the preceding section we have derived the relation (\ref{modeOP}) between the quantum-corrected
mode function and the canonical formalism. This also indicates that the expectation value of the commutator
of the field and the creation operator has the same topology as the mode function \cite{MW}.
The diagrammatic expression is depicted in Fig.~\ref{Figure2}. The main purpose of this section is to
show that the mode function definition and the 2-point correlator definition share the same
topology.

The diagram for the usual definition, the spatial Fourier transform of the 2-point correlator,
looks like,
\begin{eqnarray}
\!\int\!\!d^3x e^{i\vec{k}\cdot\vec{x}}\times\Bigl(\textrm{Fig.~1}\Bigr),
\end{eqnarray}
whereas the diagram for the $\Delta u(t,k)u^{*}(t,k)$\footnote{One can see that the phase terms
in equation (\ref{Phi0}) and (\ref{modeOP}) cancel out.}
part of the mode function definition is,
\begin{eqnarray}
\Bigl(\textrm{Fig.~2}\Bigr)\times u(y'^0,k)e^{i\vec{k}\cdot\vec{y'}}\times u^{*}(t,k).
\label{modediagram}
\end{eqnarray}
Note that the final three components in equation (\ref{modediagram}) can be re-written as,
\begin{eqnarray}
e^{i\vec{k}\cdot\vec{y'}}u(y'^0,k)u^{*}(t,k)\!=\!
\!\int\!\!d^3x' e^{i\vec{k}\cdot\vec{x'}}i\Delta_{\scriptscriptstyle -+}(y';x').
\end{eqnarray}
Actually each single propagator with different Schwinger-Keldysh polarity takes the form of
the linear combination of $uu^{*}$ and $u^{*}u$ multiplying by a distinct theta function.
Hence the mode function definition has the diagrammatic form,
\begin{eqnarray}
\!\int\!\!d^3x' e^{i\vec{k}\cdot\vec{x'}}\times \Bigl(\textrm{Fig.~1}\Bigr).\label{modeFunTopology}
\end{eqnarray}
Therefore we conclude that the generic diagram topology of the mode function definition is
identical to that of the correlator definition.

\section{K\"{a}llen Representation}

In this subsection we will point out the necessary conditions for K\"{a}llen representation
to hold through deriving it step by step. Although this familiar material has been covered
in a textbook of quantum field theory, the purpose here is to remind readers that the familiar
representation becomes nontrivial when applying it to FRW background in a theory without a
mass gap.

In an interacting theory the 2-point correlation function is,
\begin{eqnarray}
\Bigl<\Omega|\phi(x)\phi(y)|\Omega\Bigr>.\label{correlator}
\end{eqnarray}
It is free to insert a partition of unity between the two field operators,
\begin{eqnarray}
\textrm{I}=|\Omega\Bigr>\Bigl<\Omega|+\!\!\int\!\!\frac{d^3k}{(2\pi)^3}\frac{1}{2\omega}
|k\Bigr>\Bigl<k|+\!\!\int\!\!\frac{d^3k_1}{(2\pi)^3}\frac{1}{2\omega_1}
\!\!\int\!\!\frac{d^3k_2}{(2\pi)^3}\frac{1}{2\omega_2}|k_1k_2\Bigr>\Bigl<k_1k_2|+\cdots.\,
\end{eqnarray}
The 2-point correlation function (\ref{correlator}) can be written as ,
\begin{eqnarray}
&&\Bigl<\Omega|\phi(x)\phi(y)|\Omega\Bigr>=\Bigl<\Omega|\phi(x)|\Omega\Bigr>
\Bigl<\Omega|\phi(y)|\Omega\Bigr>+\!\!\int\!\!\frac{d^3k}{(2\pi)^3}\frac{1}{2\omega}
\Bigl<\Omega|\phi(x)|k\Bigr>\Bigl<k|\phi(y)|\Omega\Bigr>\nonumber\\
&&\hspace{2.5cm}+\!\!\int\!\!\frac{d^3k_1}{(2\pi)^3}\frac{1}{2\omega_1}
\!\!\int\!\!\frac{d^3k_2}{(2\pi)^3}\frac{1}{2\omega_2}
\Bigl<\Omega|\phi(x)|k_1k_2\Bigr>\Bigl<k_1k_2|\phi(y)|\Omega\Bigr>+\cdots.
\label{correlator1}
\end{eqnarray}
The first term of (\ref{correlator1}) can be subtracted from $\phi$ so the VEV of it
is zero if the quantum field has been properly defined,
\begin{eqnarray}
\Bigl<\Omega|\phi(x)|\Omega\Bigr>=\phi_0,
\end{eqnarray}
and the resulting sum begins with the contribution from 1-particle states.
Had a theory possessed poincar\'{e} symmetry, we can implement the following steps,
\begin{eqnarray}
\Bigl<\Omega|\phi(x)|k\Bigr>=e^{-ik^{\mu}x_{\mu}}\Bigl<\Omega|\phi(0)|k\Bigr>
=e^{-ik^{\mu}x_{\mu}}\Bigl<\Omega|\phi(0)|\vec{0}\Bigr>\equiv
\sqrt{Z}e^{-ik^{\mu}x_{\mu}}.\label{keyStep }
\end{eqnarray}
The first equality is from the spacetime\footnote{The metric convention here is
$+---$ rather than $-+++$ in general relativity or cosmology.} translation invariance of
the 3-momentum state $|k\Bigr>$ and the vacuum $\Bigl<\Omega|$ whereas
the second equality is due to the Lorentz boost invariance of $\phi(0)$ and
the vacuum $\Bigl<\Omega|$. This is not a problem at all for a quantum field theory
with a mass gap in a flat background. The first nonzero term of (\ref{correlator1}) is,
\begin{eqnarray}
\int\!\!\frac{d^3k}{(2\pi)^3}\frac{1}{2\omega}Ze^{-ik^{\mu}(x-y)_{\mu}}
=\!\int\!\!\frac{d^4k}{(2\pi)^4}\frac{iZe^{-ik^{\mu}(x-y)_{\mu}}}
{k^2\!-\!m^2+i\epsilon}
\end{eqnarray}
Here we have assumed $x^{0}>y^{0}$ for convenience. Therefore we reach the familiar
expression,
\begin{eqnarray}
\Bigl<\Omega|\phi(x)\phi(y)|\Omega\Bigr>=\!\int\!\!\frac{d^4k}{(2\pi)^4}
e^{-ik^{\mu}(x-y)_{\mu}}\Biggl\{\frac{iZ}{k^2\!-\!m^2+i\epsilon}
+\sum\frac{iZ(\mu)}{k^2\!-\!\mu^2+i\epsilon}\Biggr\}.\label{correlator2}
\end{eqnarray}
The first term of (\ref{correlator2}) is the 1-particle contribution and the second
term is the multi-particle continuation.

The primordial power spectrum is computed in FRW geometry which only possesses spatial
rotation and spatial translation invariance rather than poincar\'{e} symmetry. Hence
the 2-point correlation function is not able to reach the same expression as
(\ref{correlator2}). As a result, there are no exact 1-particle states, nor even
exact energy eigenstates. Nor is the VEV of the field zero, or even constant. Furthermore
the graviton in FRW geometry experiences much more severe IR divergences than in flat background
because a rapid spacetime expansion makes IR divergence much stronger\footnote{The Bloch-Nordesick
procedure for a massless theory in flat spacetime cannot entirely cure
the IR problem in cosmology.}. So it is also impossible to take the initial state back to
a distant infinity and to define the unique vacuum as we did for a theory with a mass gap
in flat spacetime.

Our toy model is in flat space background. However, because it is massless, $\varphi^3$ theory,
the vacuum decays and it doesn't have either time translation invariance or boost invariance
\cite{GV}. It would indeed have an IR divergence if the computation had been done
in 4-momentum space. However, in position space in the Schwinger-Keldysh (in-in) formalism
there is no IR divergence as long as the state is released at a finite time. What we would instead
get is a growth of the VEV of the field \cite{TW3}.

Based on the arguments being discussed in the preceding paragraphs, the mode function definition
cannot recover the 1-particle states in k\"{a}llen representation either for quantum gravity
in cosmology or for our toy model in flat spacetime. We also can see this from the diagram topology.
The 1-particle contribution has infinite corrections from summing up a series of the 2-point
correlator with more and more 1PI insertions as being shown in equation (7.43) of \cite{Peskin}.
The topology of the mode function definition has been derived in equation (\ref{modeFunTopology})
which shares the same topology as the spatial Fourier transform of the 2-point correlator.
The two definitions both do not receive infinite corrections as the 1-particle states do.
If the theory is poincar\'{e} invariant and has a mass gap, then we would get the same answer in
both the in-out and in-in formalisms by taking the initial time to minus infinity. However,
the subtle points the two definitions disagree at loop orders are that when there are particle
productions (which there is for cosmology and for massless, $\varphi^3$ even in a flat space)
and when we cannot take the initial time to minus infinity.

\section{The issues of divergences}

In this subsection we renormalize the ultraviolet divergences of the two definitions in our toy
model using a mass counterterm. For further clarification of the renormalized results we have
emphasized the distinction between infrared divergences and secular growth. Finally, we discuss
the extra, composite operator divergence which can occur in a model with derivative interactions,
owing to the fact that the two times coincide.

\subsection{The power spectrum from the mode function definition for $\varphi^3$ theory}

Before computing the lowest order correction to the power spectrum from the mode function definition
(\ref{modedef}), we need to obtain the first order correction to the mode function using (\ref{u1eqn}).
Instead of employing the formal expression (\ref{u1sol}) and (\ref{Duu3}), the best way to get
the finite result is to remove the ultraviolet divergence of the self-mass squared and then
integrate the finite part against the tree order mode function. Even though the last step to solve for
$\Delta u(t,k)$ from (\ref{u1eqn}) still requires one more integral coming from the retarded Green's function,
it is actually not so hard to perform because it only involves with a temporal integration.

Recall that the primitive part of the one loop self-mass squared in $\varphi^3$ theory is,
\begin{eqnarray}
-iM^2_{\scriptscriptstyle +\pm}(x;x')=\mp\frac{\lambda^2}{2}
\frac{\Gamma^2(\frac{D}{2}\!-\!1)}{16\pi^{D}}
\frac{1}{\Delta x_{\scriptscriptstyle +\pm}^{2D-4}(x;x')},\label{Msquare}
\end{eqnarray}
where $\Delta x^2_{\scriptscriptstyle +\pm}(x;x')$ is defined in (\ref{x++}) and (\ref{x+-}). Note that
integrating expression (\ref{Msquare}) with respect to $x^{\prime \mu}$ in $D=4$ dimensions would
produce a logarithmic divergence due to the singularity at $x^{\prime \mu} = x^{\mu}$. We can make the
expression integrable by extracting a d'Alembertian with respect to $x^{\mu}$,
\begin{eqnarray}
\int\!\!d^4x'\frac{1}{\Delta x_{\scriptscriptstyle +\pm}^{2D-4}(x;x')}
=\frac{1}{2(D\!-\!3)(D\!-\!4)}\partial^2\!\!\int\!\! d^4x'
\frac{1}{\Delta x_{\scriptscriptstyle +\pm}^{2D-6}(x;x')}.\label{extract}
\end{eqnarray}
The remaining obstacle to taking the $D \rightarrow 4$ limit is of course the explicit factor of
$1/(D-4)$ in expression (\ref{extract}).

The next step is to segregate the divergence into a local delta function by adding zero in the form,
\begin{eqnarray}
\partial^{2} \Bigl[\frac{1}{\Delta x^{D-2}_{\scriptscriptstyle ++}} \Bigr] - \frac{i 4
\pi^{\frac{D}{2}} \delta^D(x \!-\! x')}{\Gamma(\frac{D}{2}\!-\!1)} = 0 =
\partial^{2} \Bigl[\frac{1}{\Delta x^{D-2}_{\scriptscriptstyle +-}} \Bigr] \label{0}.
\end{eqnarray}
(For simplicity, we here and henceforth suppress the two arguments of the coordinate separation
$\Delta x^2(x;x')$.) We can then take the $D \rightarrow 4$ limit of the nonlocal part,
leaving the divergence restricted to the delta function. For the $++$ case the result is,
\begin{eqnarray}
\lefteqn{\frac{1}{(D\!-\!4)} \Biggl\{ \partial^2 \Bigl[\frac{1}{\Delta x^{2D-6}_{
\scriptscriptstyle ++}} - \frac{\mu^{2D-4}}{\Delta x^{D-2}_{\scriptscriptstyle ++}}
\Bigr] + \frac{\mu^{D-4} i 4 \pi^{\frac{D}{2}}}{\Gamma(\frac{D}{2}\!-\!1)}
\delta^D(x\!-\!x') \Biggr\} } \nonumber \\
& & = \frac{\mu^{D-4}}{(D\!-\!4)}\frac{i4\pi^{\frac{D}{2}}}{\Gamma(\frac{D}{2}\!-\!1)}
\delta^D(x\!-\!x') -\frac{\partial^2}{2} \Biggl[\frac{\ln(\mu^2\Delta x^2_{\scriptscriptstyle ++})}
{\Delta x^2_{\scriptscriptstyle ++}}\Biggr] + O(D \!-\! 4) . \label{segregate}
\end{eqnarray}
At this point it is clear that the divergent part of the $++$ self-mass squared is,
\begin{eqnarray}
\frac{-i\lambda^2}{2^4\pi^{\frac{D}{2}}}\frac{\Gamma(\frac{D}{2}\!-\!1)}{(D\!-\!3)}
\frac{\mu^{D-4}}{(D\!-\!4)}\delta^D(x\!-\!x'),\label{massCTT}
\end{eqnarray}
and it can be absorbed by a mass counterterm.\footnote{Recall that $\lambda$ has the dimension
of mass in $\varphi^3$ theory so mass is not multiplicatively renormalized.} Because there is no
delta function for the $+-$ term in expression (\ref{0}), the $+-$ self-mass squared has no
ultraviolet divergence. This accords with the fact that the Schwinger-Keldysh formalism has no
counterterms with mixed $\pm$ polarities \cite{CSHY,RDJ,FW}.

It is simpler to perform the integral (\ref{u1eqn}) by extracting one more d'Alembertian,
\begin{eqnarray}
\Biggl[\frac{\ln(\mu^2\Delta x^2_{\scriptscriptstyle +\pm})}{\Delta x^2_{\scriptscriptstyle +\pm}}\Biggr]
=\frac{\partial^2}{8}\Biggl\{\ln^2(\mu^2\Delta x^2_{\scriptscriptstyle +\pm})-
2\ln(\mu^2\Delta x^2_{\scriptscriptstyle +\pm})\Biggr\} . \label{extract1}
\end{eqnarray}
With the two simplifications,
\begin{eqnarray}
\ln(\mu^2\Delta x^2_{\scriptscriptstyle +\pm})=\theta(t\!-\!t')\theta(\Delta t\!-\!\Delta \overline{x})
\Biggl\{\ln[\mu^2(\Delta t^2\!-\!\Delta \overline{x}^2)]\pm i\pi\Biggr\} ,
\end{eqnarray}
the finite part can be written as,
\begin{eqnarray}
M^2_{\scriptscriptstyle ++}+M^2_{\scriptscriptstyle +-}=\frac{-\lambda^2}{2^8\pi^3}\partial^4\Biggl\{
\theta(t\!-\!t')\theta(\Delta t\!-\!\Delta \overline{x})\Biggl(
\ln[\mu^2(\Delta t^2\!-\!\Delta \overline{x}^2)]-1\Biggr)\Biggr\}.\label{massfinite}
\end{eqnarray}
Here $\Delta t^2$ and $\Delta \overline{x}^2$ are defined as $(t-t')^2$ and $||\vec{x}-\vec{x'}||^2$ respectively.

At this stage we are ready to integrate (\ref{massfinite}) against the tree order mode function,
\begin{eqnarray}
&&\int\!\!d^4x'\Bigl[M^2_{\scriptscriptstyle ++}(x;x')+M^2_{\scriptscriptstyle +-}(x;x')\Bigr]
u(t',k)e^{i\vec{k}\cdot\vec{x'}}\nonumber\\
&&\hspace{-.5cm}=\frac{-\lambda^{2}e^{i\vec{k}\cdot\vec{x}}}{2^8\pi^3}
\partial^4\!\!\int_0^t\!dt'\!\!\int_0^{\Delta t}\!\!drr^2d\Omega
\Biggl\{\ln[\mu^2(\Delta t^2\!-\!\Delta \overline{x}^2)]-1\Biggr\}
\frac{e^{-ikt'+i\vec{k}\cdot\vec{r}}}{\sqrt{2k}}.
\end{eqnarray}
Here we set $\vec{r}$ as $\vec{x'}-\vec{x}$. After executing the angular integration and change
the variable $r=|\vec{r}|=z\Delta t$, the expression can be simplified,
\begin{eqnarray}
\hspace{-1cm}\frac{-\lambda^2}{2^6\pi^2}\frac{e^{i\vec{k}\cdot\vec{x}}}{k}
(\partial_0^2\!+\!k^2)^2\!\!\int_0^t\!\!dt'\frac{e^{-ikt'}}{\sqrt{2k}}\Delta t^2\!\!
\int_0^1\!\!\!dzz\sin(kz\Delta t)
\Biggl\{2\ln(\mu\Delta t)\!+\!\ln(1\!\!-\!z^2)\!-\!1\Biggr\}\!.
\end{eqnarray}
To perform the $z$ integration we employ several special functions,
\begin{eqnarray}
&&\textrm{Si}(x)\equiv-\!\!\int_{x}^{\infty}\!\!dt\frac{\sin(t)}{t}=
-\frac{\pi}{2}+\!\!\int_0^x\!\!dt\frac{\sin(t)}{t};\nonumber\\
&&\textrm{Ci}(x)\equiv-\!\!\int_x^{\infty}\!\!dt\frac{\cos(t)}{t}=\gamma+\ln(x)
+\!\!\int_0^x\!\!dt\Biggl[\frac{\cos(t)\!-\!1}{t}\Biggr];\nonumber\\
&&\hspace{-.5cm}\xi(\alpha)\equiv\!\!\int_0^1\!dzz\sin(\alpha z)\ln(1\!-\!z^2)
=\frac{1}{\alpha^2}\Biggl\{2\sin(\alpha)\!-\!\Bigl[\cos(\alpha)\!+\!\alpha\sin(\alpha)\Bigr]
\Bigl[\textrm{Si}(2\alpha)\!+\!\frac{\pi}{2}\Bigr]\nonumber\\
&&\hspace{5cm}+\Bigl[\sin(\alpha)\!-\!\alpha\cos(\alpha)\Bigr]\Bigl[\textrm{Ci}(2\alpha)\!-\!
\gamma\!-\!\ln(\frac{\alpha}{2})\Bigr]\Biggr\}.
\end{eqnarray}
With these the renormalized result can be expressed as,
\begin{eqnarray}
\hspace{-.3cm}\frac{-\lambda^2}{2^6\pi^2}\frac{e^{i\vec{k}\cdot\vec{x}}}{k^3}
(\partial_0^2\!+\!k^2)^2\!\!\int_0^t\!\!dt'\frac{e^{-ikt'}}{\sqrt{2k}}
\Biggl\{\!\alpha^2\xi(\alpha)\!+\!\Bigl[2\ln(\frac{\mu\alpha}{k})\!-\!1\Bigr]
\Bigl[\sin(\alpha)\!-\!\alpha\cos(\alpha)\Bigr]\!\Biggr\}\!. \label{xiform}
\end{eqnarray}
Here $\alpha$ is $k\Delta t$.

Because the integrand of (\ref{xiform}) behaves like $\Delta t^3\ln(\Delta t)$ near
$t'\!=\!t$ we can pass three of four derivatives through the integral sign to simplify
the integrand. Passing the first two derivatives through gives,
\begin{eqnarray}
&&\hspace{-1.7cm}\frac{-\lambda^2}{2^5\pi^2}\frac{e^{i\vec{k}\cdot\vec{x}}}{k}
(\partial_0\!+\!ik)(\partial_0\!-\!ik)\!\!\int_0^t\!\!dt'\frac{e^{-ikt'}}{\sqrt{2k}}
\Biggl\{\!-\cos(\alpha)\!\!\int_0^{2\alpha}\!\!ds\frac{\sin(s)}{s}\nonumber\\
&&\hspace{3.7cm}+\sin(\alpha)\Biggl[\int_0^{2\alpha}\!\!ds\frac{\cos(s)\!-\!1}{s}
\!+\!2\ln(\frac{2\mu\alpha}{k})\Biggr]\Biggr\}.
\end{eqnarray}
Extracting the temporal phase factor and passing one more derivative through the
integral gives,
\begin{eqnarray}
&&\hspace{-1.3cm}\frac{-\lambda^2}{2^5\pi^2}e^{i\vec{k}\cdot\vec{x}}(\partial_0\!+\!ik)
\!\!\int_0^t\!\!dt'\frac{e^{-ikt'}}{\sqrt{2k}}e^{-ik\Delta t}\Biggl[
\!\int_0^{2\alpha}\!\!ds\frac{e^{is}-1}{s}\!+\!2\ln(2\mu\Delta t)\Biggr]\nonumber\\
&&\hspace{-.5cm}=\frac{-\lambda^2}{2^5\pi^2}\frac{e^{-ikt+i\vec{k}\cdot\vec{x}}}{\sqrt{2k}}
\partial_0\!\int_0^t\!\!d\Delta t\times 1\times\Biggl\{\Biggl[\!\int_0^{2k\Delta t}\!\!ds
\frac{e^{is}-1}{s}\Biggr]\!+\!2\ln(2\mu\Delta t)\Biggr\}.
\end{eqnarray}
Here we have used $(\partial_0+ik)e^{-ikt} = 0$. Further simplification can be accomplished
by performing the $\Delta t$ integration and acting the final derivative. The final
result is,
\begin{eqnarray}
\frac{-\lambda^2}{2^5\pi^2}\frac{e^{-ikt+i\vec{k}\cdot\vec{x}}}{\sqrt{2k}}
\Biggl\{\!\int_0^{2kt}\!\!ds\frac{e^{is}-1}{s}\!+\!2\ln(2\mu t)\Biggr\}
\equiv-S(t)e^{i\vec{k}\cdot\vec{x}}.
\end{eqnarray}

According to (\ref{u1eqn}) the $\Delta u(t,k)$ we want to solve for obeys,
\begin{eqnarray}
&&\hspace{-1.5cm}\mathcal{D}\Bigl[\Delta u(t,k)e^{i\vec{k}\cdot\vec{x}}\Bigr]\!=\!
-(\partial_0^2\!+\!k^2)\Delta u(t,k)e^{i\vec{k}\cdot\vec{x}}\!=\!
-S(t)e^{i\vec{k}\cdot\vec{x}}\,\,\,\Longrightarrow\nonumber\\
&&(\partial_0^2\!+\!k^2)\Delta u(t,k)\!=\!S(t)\,\,\Longrightarrow\,\,
\Delta u(t,k)\!=\!\!\int_0^{\infty}\!\!dt'G_r(t,t')S(t').\label{Deltau}
\end{eqnarray}
Here $G_r(t,t')=\theta(t\!-\!t')\frac{\sin(k\Delta t)}{k}$ is the retarded Green's function.
Plugging the explicit forms of $G_r(t,t')$ and $S(t')$ into (\ref{Deltau}) gives,
\begin{eqnarray}
\Delta u(t,k)\!=\!\frac{\lambda^2}{2^5\pi^2}\frac{e^{-ikt}}{i(2k)^{\frac32}}
\!\int_0^t\!\!dt'\Bigl\{e^{2ik\Delta t}\!-\!1\Bigr\}\Biggl\{\!\int_0^{2kt'}\!\!ds
\frac{e^{is}-1}{s}\!+\!2\ln(2\mu t')\Biggr\}.
\end{eqnarray}
It remains to perform the four $t'$ integrations and collect terms. The result is,
\begin{eqnarray}
&&\hspace{-2cm}\Delta u(t,k)\!=\!\frac{\lambda^2}{2^5\pi^2}\frac{e^{-ikt}}{(2k)^{\frac32}}\Biggl\{\!
\Biggl[\frac1{2k}\!+\!it\Biggr]\!\int_0^{2kt}\!\!ds\frac{e^{is}\!-\!1}{s}\!-\!\frac{e^{2ikt}}{2k}
\!\!\int_0^{2kt}\!\!ds\frac{e^{-is}\!-\!1}{s}\nonumber\\
&&\hspace{3cm}+\Biggl[\frac{1\!-\!e^{2ikt}}{k}\!+\!2it\Biggr]\ln(2\mu t)\!-\!it\!+\!
\frac{1\!-\!e^{2ikt}}{2k}\Biggr\}.\label{FnlDeltau}
\end{eqnarray}
Combining (\ref{FnlDeltau})$\times [u^*(t,k)=\frac{e^{-ikt}}{\sqrt{2k}}]$ with
its complex conjugate gives the lowest-order correction to the power spectrum,
\begin{eqnarray}
&&\hspace{-1.5cm}\Delta u(t,k)u^*(t,k)\!+\!\textrm{c.c.}\!=\!\frac{\lambda^2}{2^7\pi^2}
\frac1{k^3}\Biggl\{\Bigl[1\!-\!\cos(2kt)\Bigr]\Bigl[1\!-\!\gamma\!+\!\textrm{Ci}(2kt)
\!+\!\ln(\frac{2\mu^2t}{k})\Bigr]\nonumber\\
&&\hspace{5cm}-\Bigl[\sin(2kt)\!+\!2kt\Bigr]\Bigl[\frac{\pi}{2}
\!+\!\textrm{Si}(2kt)\Bigr]\Biggr\}.\label{reMoSpectra}
\end{eqnarray}

\subsection{The power spectrum from the correlator definition for $\varphi^3$ theory}

We begin by removing the ultraviolet divergence of the self-mass squared embedded
in the 2-point correlator using the same procedure prescribed in the previous subsection.
We also perform two partial integrations and carry out the spatial Fourier transforms.

Even though the 2-point correlator carries various polarities on the self-mass squared
and the external legs (Fig.\ref{Figure1}), we suppress the polarities and the relative
signs\footnote{We used the convention for the usual, in-out diagram.} in order to
investigate the generic pattern ,
\begin{eqnarray}
&&\hspace{-1.5cm}\frac{-\lambda^2}{2}\Biggl[\frac{\Gamma(\frac{D}{2}\!-\!1)}{4\pi^{\frac{D}{2}}}\Biggr]^4
\!\!\int\!\!d^{D}y\frac{1}{\Delta x^{\scriptscriptstyle D-2}(x;y)}\!\!\int\!\!d^{D}y'
\frac{1}{\Delta x^{\scriptscriptstyle 2D-4}(y;y')}\frac{1}
{\Delta x^{\scriptscriptstyle D-2}(x';y')},\nonumber\\
&&\hspace{-1.5cm}\Longrightarrow\frac{-\lambda^2}{2}\frac{1}{2^8\pi^8}\!\!\int\!\!d^{4}y
\frac{1}{\Delta x^{\scriptscriptstyle 2}(x;y)}\!\!\times\!\!\frac{-1}{4}\!\!\int\!\!d^{4}y'
\partial_y^2\Biggl[\frac{\ln[\mu^{\scriptscriptstyle 2}\Delta x^{\scriptscriptstyle 2}(y;y')]}
{\Delta x^{\scriptscriptstyle 2}(y;y')}\Biggr]\frac{1}{\Delta x^{\scriptscriptstyle 2}(x';y')}.
\end{eqnarray}
To reach the second line, we have employed (\ref{extract}) to make the function integrable
with respect to $y^{\prime \mu}$ in $D=4$ dimensions. We then segregated the ultraviolet
divergence into a local delta function using (\ref{0}) and absorbed it with a mass counterterm,
just as in the previous subsection.

Because the derivative only acts on a function of the coordinate separation, we can replace
$\partial^2_y$ with $\partial^2_{y'}$ and then partially integrate to reach the form,
\begin{eqnarray}
\frac{\ln[\mu^{\scriptscriptstyle 2}\Delta x^{\scriptscriptstyle 2}(y;y')]}
{\Delta x^{\scriptscriptstyle 2}(y;y')}\Biggl[\partial_y'^2
\frac{1}{\Delta x^{\scriptscriptstyle 2}(x';y')}\Biggr]\!+\!\textrm{two surface terms}.
\end{eqnarray}
After dropping the surface terms, the $y'$ integration can be carried out using (\ref{0}),
\begin{eqnarray}
\frac{i\lambda^2}{2^9\pi^6}\!\!\int\!\!d^4y\frac{1}{\Delta x^{\scriptscriptstyle 2}(x;y)}
\frac{\ln[\mu^{\scriptscriptstyle 2}\Delta x^{\scriptscriptstyle 2}(y;x')]}
{\Delta x^{\scriptscriptstyle 2}(y;x')}.
\end{eqnarray}
Note that only the first and third diagrams of Fig.\ref{Figure1} survive because their
right external legs carry the same polarity.

Further reduction can be accomplished by extracting another d'Alembertian using (\ref{extract1}),
performing a partial integration and then carrying out the $y$ integration,
\begin{eqnarray}
\frac{\lambda^2}{2^{10}\pi^4}\Biggl\{\ln^2[\mu^{\scriptscriptstyle 2}
\Delta x^{\scriptscriptstyle 2}_{\scriptscriptstyle -+}(x;x')]
\!-\!2\ln(\mu^{\scriptscriptstyle 2}
\Delta x^{\scriptscriptstyle 2}_{\scriptscriptstyle -+}(x;x')]
\Biggr\}\!+\!\textrm{two surface terms}. \label{re2pt}
\end{eqnarray}
It is clear that the final survival term is from the third diagram of Fig.\ref{Figure1}
because its two external legs carry the same polarity. Because $x^{\mu} \!=\!(t,\vec{x})$
and $x^{\prime \mu} \!=\!(t,\vec{0})$ have the same time components, the coordinate separation
$\Delta x^2(x;x')$ in (\ref{re2pt}) is purely spatial. The spatial Fourier transform of
(\ref{re2pt}) can be easily performed to give,
\begin{eqnarray}
&&\hspace{-1cm}\frac{\lambda^2}{2^{10}\pi^4}\!\!\int\!\!d^3x e^{-i\vec{k\cdot\vec{x}}}
\Biggl\{\!\ln^2[\mu^{\scriptscriptstyle 2}|\vec{x}|^{\scriptscriptstyle 2}]
\!-\!2\ln[\mu^{\scriptscriptstyle 2}|\vec{x}|^{\scriptscriptstyle 2}]\!\Biggr\}
\!=\!\frac{\lambda^2}{2^{10}\pi^4}\frac{4^2\pi}{k}\!\!\int_0^{\infty}\!\!drr\sin(kr)
\Big\{\!\ln^2[\mu r]\!-\!\ln[\mu r]\!\Bigr\}\nonumber\\
&&\hspace{-.7cm}=\frac{\lambda^2k^{-1}}{2^{6}\pi^3}\frac{-\partial}{\partial k}\!\!\int_0^{\infty}\!\!
dr\cos(kr)\Biggl\{\ln^2[\mu r]\!-\!\ln[\mu r]\Biggr\}\!=\!\frac{\lambda^2k^{-1}}{2^6\pi^3}
\frac{-\partial}{\partial k}\Biggl\{\frac{\pi}{k}\Bigl[\frac12\!+\!\gamma\!-\!
\ln(\frac{\mu}{k})\Bigr]\Biggr\}.
\end{eqnarray}
Two special integrals \cite{integrals} have been employed in the last equality,
\begin{eqnarray}
\int_0^{\infty}\!\!dz\frac{\sin(z)}{z}\!=\!\frac{\pi}{2}\,\,\,;\,\,\,\!\!\int_0^{\infty}\!\!
dz\frac{\sin(z)}{z}\ln(z)\!=\!\frac{\pi}{2}\gamma,\,\,\,\,\,\gamma\equiv \textrm{Euler's constant}.
\end{eqnarray}
The lowest correction to the power spectrum by the correlator definition is therefore,
\begin{eqnarray}
\int\!\!d^3x e^{-i\vec{k\cdot\vec{x}}}
\Bigl<\Omega|\varphi(t,\vec{x})\varphi(t,\vec{0})|\Omega\Bigr>_{\scriptstyle 1\,loop}
\!=\!\frac{\lambda^2}{2^6\pi^2}\frac{1}{k^3}\Bigl\{\frac12\!+\!\gamma\!-\!
\ln(\frac{\mu}{k})\Bigr\}.\label{2ptSpectra}
\end{eqnarray}

\subsection{Discussions of infrared and ultraviolet divergences}

There is an unfortunate tendency in the literature to employ the term ``infrared divergence''
to describe perfectly finite, temporally growing effects such as (\ref{reMoSpectra}). Of
course a true infrared divergence is an infinite constant, with no spacetime dependence.
That neither definition for the power spectrum can give rise to an infrared divergence is a
simple consequence of the Schwinger-Keldysh formalism with the initial states being released
at finite times. In order to produce a true infrared divergence, interactions must contribute
from arbitrarily large spatial distances, and this is precluded by causality as long as the
initial state is released at any finite time.

Ultraviolet divergences can and do occur in the Schwinger-Keldysh formalism, just as they
do in the in-out formalism. In both formalisms it is important to distinguish between the
ultraviolet divergences of non-coincident 1PI functions, which are eliminated by
conventional BPHZ renormalization, and the new divergences which can occur when one or
more of the coordinates are related. These new divergences require an extra, composite
operator renormalization. The case of the power spectrum is especially tricky because
only the time components of the two spacetime points are made to coincide. For the case
of our toy $\varphi^3$ model, this produces no extra divergence. The vertices of quantum
gravity contain derivatives, which increases the tendency for divergences. Frob, Roura
and Verdaguer have claimed that this is enough to cause the one loop correction to the
correlator definition of the power spectrum to harbor a new, composite operator
divergence. One of our points is that the mode function definition is free from this
new divergence.

\end{appendices}

\end{document}